%% file: templateArxiv.tex
\documentclass{article}

\usepackage{PRIMEarxiv}

\usepackage[utf8]{inputenc} 
\usepackage[T1]{fontenc}    
\usepackage{hyperref}       
\usepackage{url}            
\usepackage{booktabs}       
\usepackage{amsfonts}       
\usepackage{nicefrac}       
\usepackage{microtype}      
\usepackage{lipsum}
\usepackage{amsmath}        
\usepackage{subfigure}
\usepackage{fancyhdr}       
\usepackage{graphicx}       
\graphicspath{{Figures/}}     

\title{Conditional Image Prior for Uncertainty Quantification in Full Waveform Inversion 
}

\author{
  Lingyun Yang[1],\and
  Omar M. Saad[2],\and
  Guochen Wu[1],\and
  Tariq Alkhalifah[2]\and
  {[1] China University of Petroleum (East China)}\\
  {[2] King Abdullah University of Science and Technology}\\
 {\{yang15712721245@126.com; guochenwu@upc.edu.cn}\\
 {\{Omar.sadaly,tariq.alkhalifah\}@kaust.edu.sa}\\
}

\begin{document}
\maketitle

\begin{abstract}
Full Waveform Inversion (FWI) is a technique employed to attain a high resolution subsurface velocity model. However, FWI results are effected by the limited illumination of the model domain and the quality of that illumination, which is related to the quality of the data. Additionally, the high computational cost of FWI, compounded by the high dimensional nature of the model space, complicates the evaluation of model uncertainties. Recent work on applying neural networks to represent the velocity model for FWI demonstrated the network's ability to capture the salient features of the velocity model. The question we ask here is how reliable are these features in representing the observed data contribution within the model space (the posterior distribution). To address this question, we propose leveraging a conditional Convolutional Neural Network (CNN) as image prior to quantify the neural network uncertainties. Specifically, we add to the deep image prior concept a conditional channel, enabling the generation of various models corresponding to the specified condition. We initially train the conditional CNN to learn (store) samples from the prior distribution given by Gaussian Random Fields (GRF) based perturbations of the current velocity model. Subsequently, we use FWI to update the CNN model representation of the priors so that it can generate samples from the posterior distribution. These samples can be used to measure the approximate mean and standard deviation of the posterior distribution, as well as draw samples representing the posterior distribution. We demonstrate the effectiveness of the proposed approach on the Marmousi model and in a field data application.
\end{abstract}

\input{01_Introduction}
\input{02_Theory}
\input{03_Numerical_Examples}

\input{04_Field_Data_Application}

\input{05_Discussions}
\input{06_Conclusions}
\input{07_Acknowledgement}

\bibliographystyle{unsrt}  
\bibliography{references}

\end{document}

%% file: 01_Introduction.tex
\section{Introduction}
Full Waveform Inversion (FWI) is a process of iteratively updating the velocity model by minimizing the misfit between the simulated and the observed data \cite{tarantola1984inversion, pratt1999seismic, shin2008waveform, shin2009waveform, virieux2009overview, operto2013guided}. This optimization makes full use of wavefield information (e.g., amplitude and phase) to reconstruct subsurface velocity models. However, FWI is a highly nonlinear and ill-posed problem, and thus, often suffers from cycle-skipping, yielding false solutions \cite{tarantola1982generalized, mosegaard1995monte,zhang2016robust,dellinger2017garden,assis2024investigating}. In addition, FWI's dependency on dense seismic data and accurate wave propagation modeling poses challenges in practical applications, particularly in areas with limited data coverage or complex geological structures. Thus, evaluating the uncertainty in the inverted model is essential for decision making \cite{sun2024enabling, sun2024invertible}. However, FWI is costly, with a high dimensional model space, which makes uncertainty quantification extremely expensive.

Several approaches have been proposed to assess the uncertainty of FWI, e.g., the Laplace approximation approach \cite{liu2019square, fichtner2011resolution} and the Markov chain Monte Carlo (MCMC) method \cite
{sen2017transdimensional, izzatullah2021bayesian}. The Laplace approximation is the simplest of the family of posterior approximations for high-dimensional problems. However, when applied to FWI, numerous forward and adjoint operator evaluations per model point are required to approximate its covariance matrix. Monte Carlo sampling methods provide a general way to solve nonlinear inverse problems and quantify uncertainties. Similar to the Laplace approximation, the MCMC method is also extremely expensive, and it is difficult to compute in parallel when applied to FWI. Different from these popular approaches, the variational inference (VI) has recently garnered increasing interest in the geophysical community \cite{yin2024wise}. Stein Variational Gradient Descent (SVGD), as one of the VI algorithms, has demonstrated success in conducting Bayesian uncertainty analysis of FWI \cite{zhang2021bayesian, zhang20233}. Nevertheless, SVGD still suffers from high computational costs as it requires a large number of particles to sample the posterior properly \cite{izzatullah2023frugal}.

The widespread adoption of deep learning algorithms and advancements in computing infrastructure present an opportunity for significant advancements in seismic inversion processes \cite{sun2023implicit, zhang2023multilayer, fang2024deep, cheng2024self}. The current trend in geophysical inversion involves the usage of Deep Neural Networks (DNNs) to map the data to the model directly usually learned in a supervised way \cite{yang2019deep, li2019deep, fabien2020seismic,  liu2021deep, kazei2021mapping, ovcharenko2022multi, du2022deep}. However, the generalizability of the direct supervised data-driven approaches is constrained by the size and diversity of the training set \cite{lin2023physics}. There is a lack of physics information in the network, which means the reliability and generalization ability of such methods cannot be guaranteed. In addition, the needed training datasets may not be available, especially for field data. In order to get rid of these drawbacks, nowadays scientists are seeking to combine neural networks with physical information, specifically in physics informed neural networks \cite{karimpouli2020physics, song2021wavefield, huang2022pinnup, cheng2024meta, cheng2024robust}, velocity model generation method \cite{cross2002making,  feng2021multiscale}. 

The concept of deep image prior (DIP) \cite{ulyanov2018deep} demonstrates that neural networks have the ability to learn the distribution of data, like images. Building upon this feature, a CNN-based image prior velocity model representation \cite{wu2019parametric} was updated using FWI. This approach aims to minimize the misfit between simulated and observed data by iteratively updating the parameters of the CNN  used to generate the velocity model. Similar work has been described by \cite{zhu2022integrating} and \cite{he2021reparameterized}. Many applications combine different neural networks, such as Generative Adversarial Network (GAN) and Generative diffusion models, with seismic inversion \cite{zhang2020data, jin2021unsupervised, yang2023fwigan, wang2023prior, dhara2023elastic}. These studies indicate that combining neural networks and FWI is a viable solution for building more stable inverted velocity models, as we injected our prior knowledge to FWI. 

DIP can automatically capture the salient features of the velocity model \cite{wu2019parametric}. Combined with FWI, the CNN weights can be iteratively updated by reducing the data misfit. The features captured in the CNN pertaining part can be regarded as a form of regularization. Similar to traditional FWI, we can get a more stable FWI result by representing the velocity model using a neural network. However, we need to ensure whether the features captured by the neural network are reliable. 

In this paper, we propose to add a condition to the CNN image prior to store a population of velocity models as a representation of the prior distribution, with each velocity model identified by a condition value. Specifically, we construct the prior by adding Gaussian Random Fields (GRF) perturbations to the initial or current velocity model. Subsequently, the pre-trained model is integrated into the FWI process to update its weight so that it is able after convergence to generate samples representing the posterior distribution. Finally, we obtain the uncertainty map by computing the mean and standard deviation of the generated samples. Differentiating from traditional FWI, which primarily focuses on uncertainty about the inverted model, our approach instead assesses the uncertainty of the features captured by the conditional CNN. The contributions of this paper can be summarized as follows:
\begin{itemize}
\item We propose to incorporate a condition into the deep image prior (DIP) to store the prior distribution of velocity models.
\item Using FWI, we iteratively update the weights of the FWI for all the stored conditions to obtain the posterior distribution.
\item We use the posterior distribution to evaluate the uncertainty of the features in the velocity model captured by DIP, as well as sample from the distribution.
\item Numerical examples demonstrate the ability of the conditional CNN in providing reliable uncertainty quantification.
\end{itemize}

This paper is organized as follows: In the theory section, we first introduce the Gaussian random fields perturbation model to construct the prior distribution, then describe the concept of conditional Deep Image prior network and its application for uncertainty quantification of FWI. In the numerical experiment section, the Marmousi model and field data application are used to demonstrate the feasibility of the proposed method. In the discussion section, we present the effects of the network size on uncertainty quantification. We, finally, summarize our results in the conclusion.\\

%% file: 02_Theory.tex
\section{Theory}
In this section, we leverage the Conditional Deep image prior (DIP) by adding a condition to the deep image prior while storing a population of velocity models. We focus on estimating the uncertainty map (i.e., relative standard deviation), which informs us of the credibility of features captured by the DIP. The proposed conditional CNN FWI uncertainty quantification mainly includes two steps: pretraining a conditional Deep image prior to capture the prior distribution and subscribing it to the FWI process.  

Given the observed data $d$, we estimate the uncertainties of an unknown velocity model $m$. Based on Bayes' Theorem, we have
    \begin{equation}
        p(m|d) \propto p(d|m)p(m),
    \end{equation}
where $p(m)$ is the prior distribution, $p(d|m)$ denotes the likelihood and $p(m|d)$ represents the posterior distribution. We will first discuss the prior disribution.

\subsection*{Gaussian Random Fields perturbation model}
\begin{figure*}[!tb]

\centering
\includegraphics[width=0.32\linewidth]{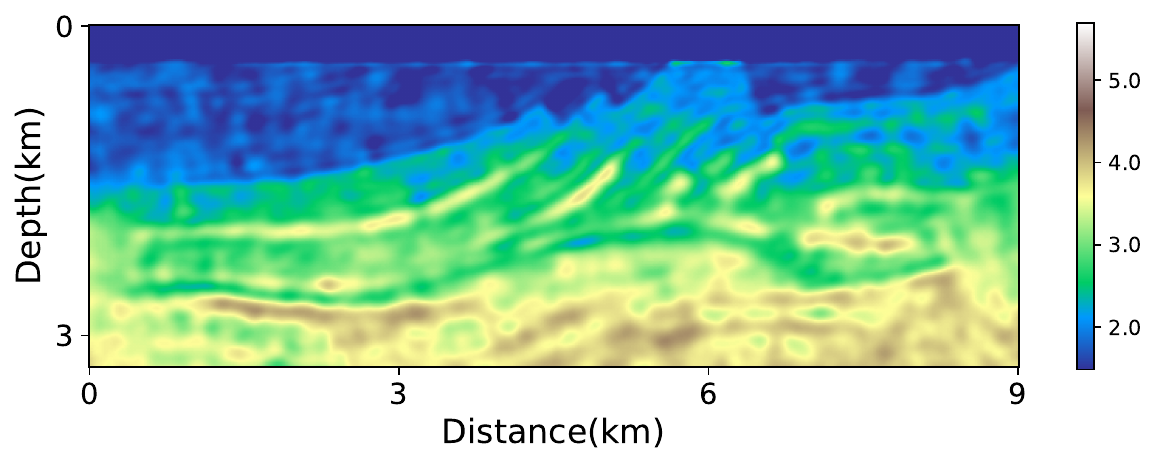}
\includegraphics[width=0.32\textwidth]{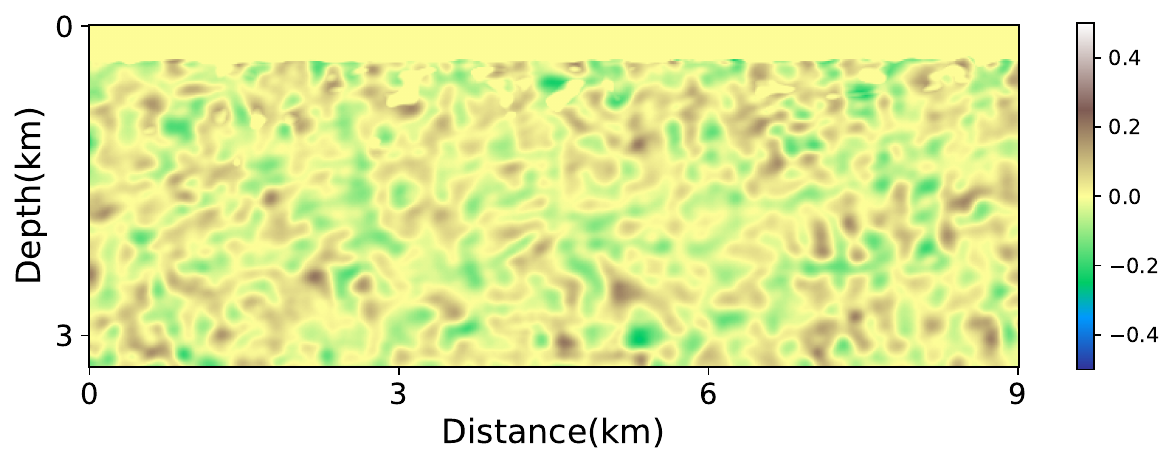}
\includegraphics[width=0.32\textwidth]{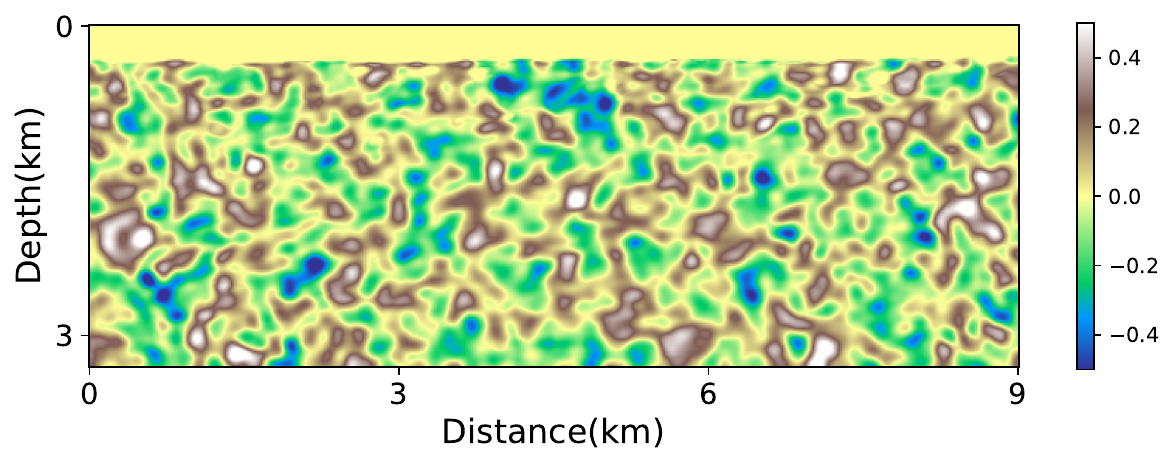}
\includegraphics[width=0.32\textwidth]{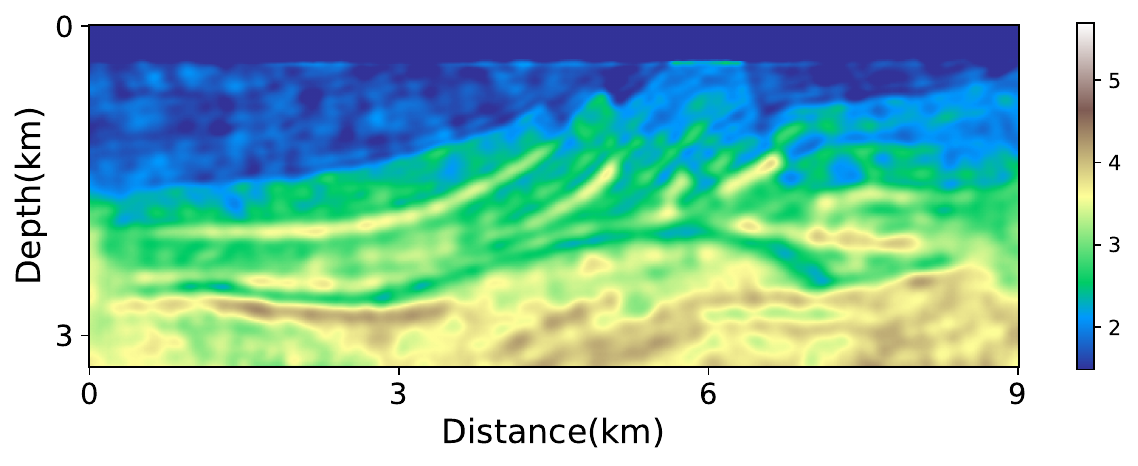}
\includegraphics[width=0.32\textwidth]{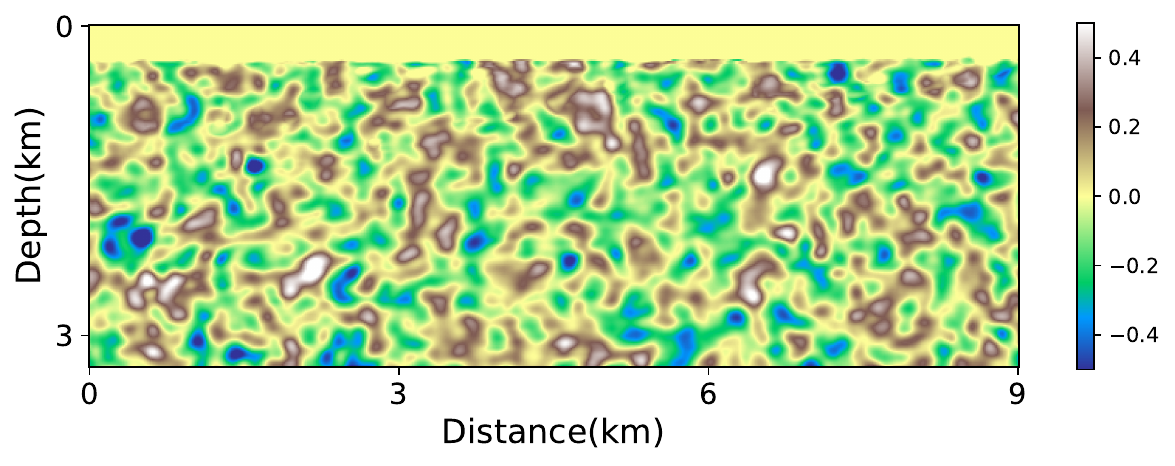}
\includegraphics[width=0.32\textwidth]{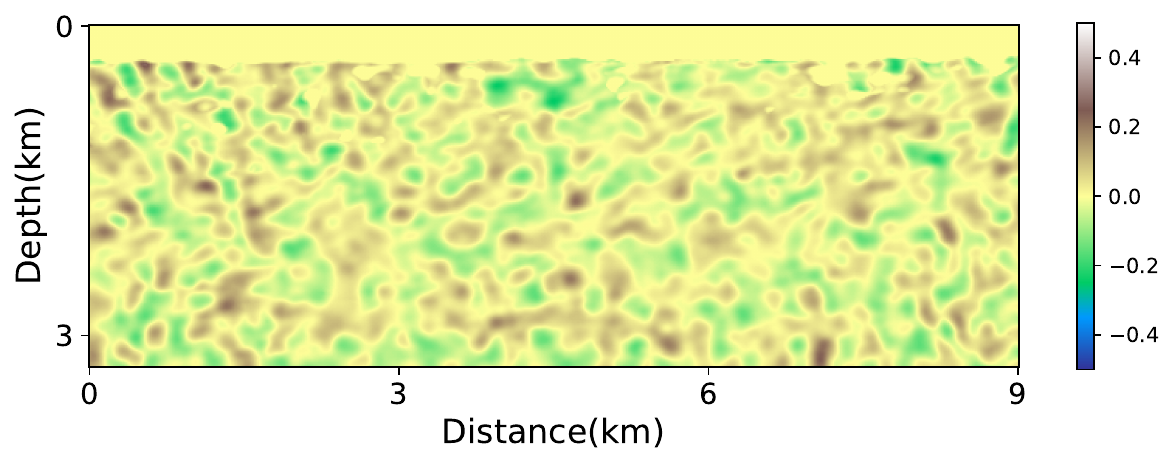}
\caption{Examples of the generated models from alternative condition vector values. The first column denotes the generated models of alternative vector values of [0.8, 0.2, ..., 0] (top row), and [0.2, 0.8, 0, ..., 0] (bottom row). The second column represents the difference between alternative vector values to the reference one-hot vector [1, 0, ..., 0]. The third column represents the difference between alternative vector values to the reference one-hot vector [0, 1, ..., 0].}
\label{fig:4-alternative-vector}
\end{figure*}

In order to capture the associated uncertainty, the prior models used in the uncertainty estimation procedure are crucial. Here, we use Gaussian Random Fields (GRF) perturbations to the initial (or current) velocity model to obtain a prior sample. The GRF perturbations provide priors that combine the propagation and scattering features of the wavefield solution \cite{izzatullah2023frugal}. We incorporate GRF perturbations with the Matern covariance kernel function $C_\eta$ \cite{bogachev1998gaussian} such that
    \begin{equation}
          C_ {\eta } (\beta)= \epsilon ^ {2} \frac {2^ {1-\eta}}{\Gamma(\eta )} ( \sqrt{2} \alpha
  \frac {\beta }{\alpha }  )  ^ {\eta }  K_{\eta }  (  \sqrt {2\eta }  \frac {\beta }{\alpha}),
    \end{equation}

where $\epsilon$ is the variance of the Gaussian process, $\Gamma$ denotes the gamma function \cite{artin2015gamma}, $K_{\eta }$ represents the modified Bessel function of the second kind, $\alpha$ is a positive parameter and $\eta$ is the smoothness parameter of the random field. The distance $\frac {\beta }{\alpha }$ between two points $m=(m_1,...,m_n)$ and $m^{'}=(m_1^{'},...,m_n^{'})$ can be calculated as follows:
\begin{equation}
    \frac {\beta }{\alpha }=\sqrt{ \sum\limits_{j=1}^{n}  (\frac{m_i-m_i^{'}}{\lambda_i})^2},
\end{equation}

where $\lambda_i$ denotes the coefficient representing the correlation length between the two points. Examples of Gaussian random fields are shown in Figure \ref{fig:4-alternative-vector}.\\ 
\begin{figure*}[!tb]
  \centering
  \includegraphics[width=\textwidth]{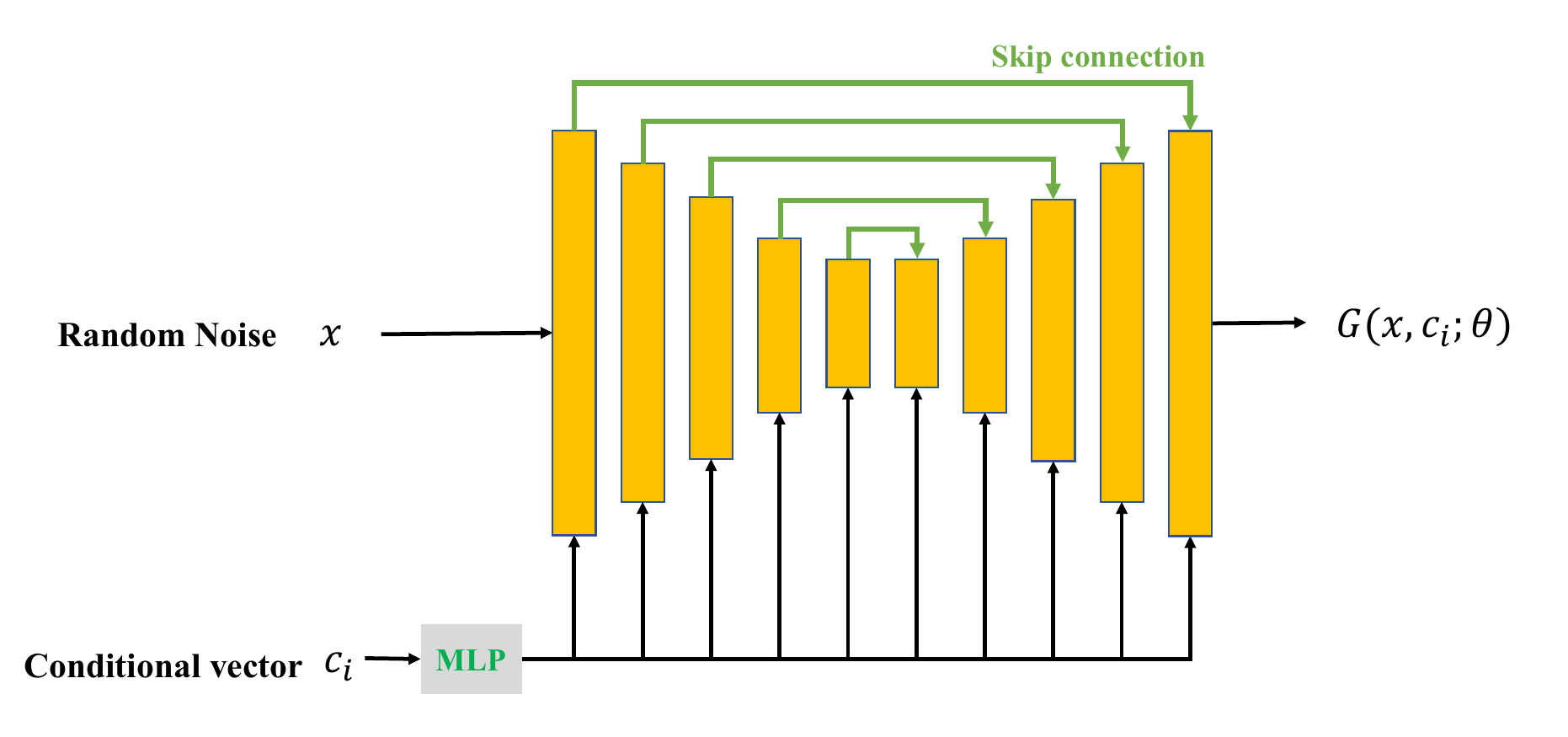}
  \caption{The architecture of the proposed conditional Convolutional Neural Network. Here, $x$ stands for the input drawn from a normal distribution, G is the generator of velocity models as a functions of $x$, $c_i$ and the network parameters $\theta$.}
  \label{fig:1-conditional-cnn-network}
\end{figure*}

\subsection*{Conditional Convolutional Neural Network}
From the concept of deep image prior (DIP), the relation between a CNN and the generated model can be expressed as
    \begin{equation}
        m=G(x;\theta),
    \end{equation}
where $m$ is the generated velocity model, $x$ is the input (the latent vector), $\theta$ represents the neural network parameters, and $G$ denotes the CNN.

In conditional DIP, two inputs are used. The first input denoted as $x$, aligns with the traditional DIP framework. The second input, referred to as a condition, guides CNN to generate diverse target models. The representation of conditional DIP can be expressed as:
    \begin{equation}
        m_{c_i}=G(x,c_i;\theta),
    \end{equation}
where $m_{c_i}$ represents the generated model corresponding to the specific condition $c_i$. Essentially, it implies that the CNN will be trained to generate the model corresponding to the input condition. Next, we will discuss the details of incorporating the condition in the CNN.

Figure \ref{fig:1-conditional-cnn-network} shows the architecture of the proposed conditional CNN network. Here, we choose the U-Net architecture, which consists of an encoder that is used to capture the geological features and a symmetric shape of a decoder that enables precise reconstruction. The conditional label ($c_i$) is first embedded as a vector and then concatenated into each hidden layer of the U-Net. The effective reconstruction of the network increases as the input goes deeper into the network, where kernel size of 3$*$3 is used. The encoder consists of 5 layers, each with feature maps of 16, 32, 64, 128, and 256. Skip connections are adopted between the encoder and decoder to combine the local and global feature maps.

The conditional label ($c_i$) stands for different input conditions, which is a one-hot vector in this study. Specifically, it is a binary vector where only one element is 1 and all others are zeros. 
For a vector of length $N$, which corresponds to $N$ different conditions (particle samples), each condition is assigned a unique index from 1 to $N$. The conditional label can be defined as:
\begin{equation}
    c_i =
    \begin{cases}
    1 & \text{if } i = k \\
    0 & \text{otherwise}
    \end{cases},
\end{equation}
where $k$ is the index of the condition, and \( i = 1, 2, ..., N \).

\begin{figure*}[!tb]
  \centering
  \includegraphics[width=\textwidth]{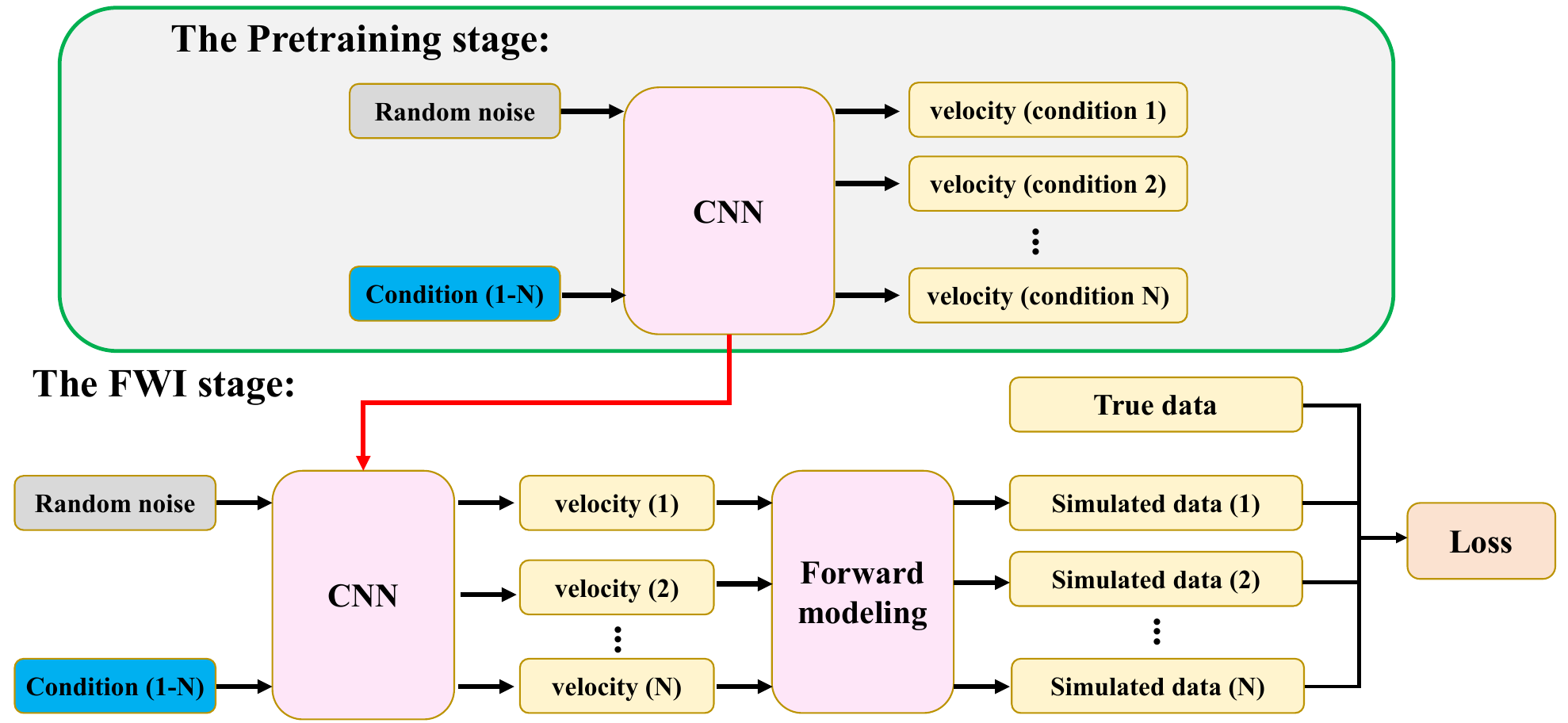}
  \caption{The proposed workflow for evaluating FWI uncertainty using deep learning. CNN here stands for the convolutional neural network, which is a Unet in this study, and N is the number of velocity models stored in the CNN.}
  \label{fig:2-workflow-of-cnn}
\end{figure*}

For the pre-training process, the conditional CNN loss function can be defined as
    \begin{equation}
        E= ||m_{t_i}-G(x,c_i;\theta)||_2^2,
    \label{equation3}
    \end{equation}
where $m_{t_i}$ is the target velocity model corresponding to the condition $c_i$. As shown in the pretraining stage of Figure \ref{fig:2-workflow-of-cnn}, the target is the velocity model with different GRFs added, where the proposed CNN network is trained in a supervised way. Since the underlying velocity is common between all models, the condition is mainly defining the added GRF.\\

\subsection*{Uncertainty quantification for FWI}
Unlike traditional FWI, where the velocity model is represented on a regular grid, the velocity model here is stored in a generator given by a neural network. Thus, the uncertainty addressed here corresponds to the network parameters as a representation of the velocity model, and the regularization involved in the salient features of the neural network. We will subscribe here to the following objective function $J$ give by the $L_2$ norm:
    \begin{equation}
        J = ||d_{true}-F(G(x,c_i;\theta))||_2^2,
    \label{equation4}
    \end{equation}
where $d_{true}$ is the observed data and $F$ denotes the wave equation operator (i.e., acoustic wave equation), mapping generated velocity model to simulated data.

Specifically, as shown in Figure \ref{fig:2-workflow-of-cnn}, the process of conditional CNN FWI inversion includes the following steps in every FWI iteration:
\begin{itemize}
    \item [1)]
     Training the conditional neural network to generate prior models corresponding to the GRF perturbations.
    \item [2)]
    Performing forward modeling to compute the corresponding synthetic data for these models. 
    \item [3)]   
    Measure the loss between observed and simulated data for the sampled velocity models based on the condition values, and subsequently use backpropagation to update the parameters of the conditional CNN.
\end{itemize}

In each FWI iteration, newly drawn random noise from a normal distribution is used as input to the CNN to mitigate the potential for overfitting of the conditional CNN. The beauty of storing the prior (and posterior) distribution in a conditional CNN network lies in its capability to instantaneously extract infinite number of samples by inputting new vectors into the condition, where the sum of its elements equals to one.\\ 
\begin{figure*}[!tb]
  \centering
    \includegraphics[width=0.32\textwidth]{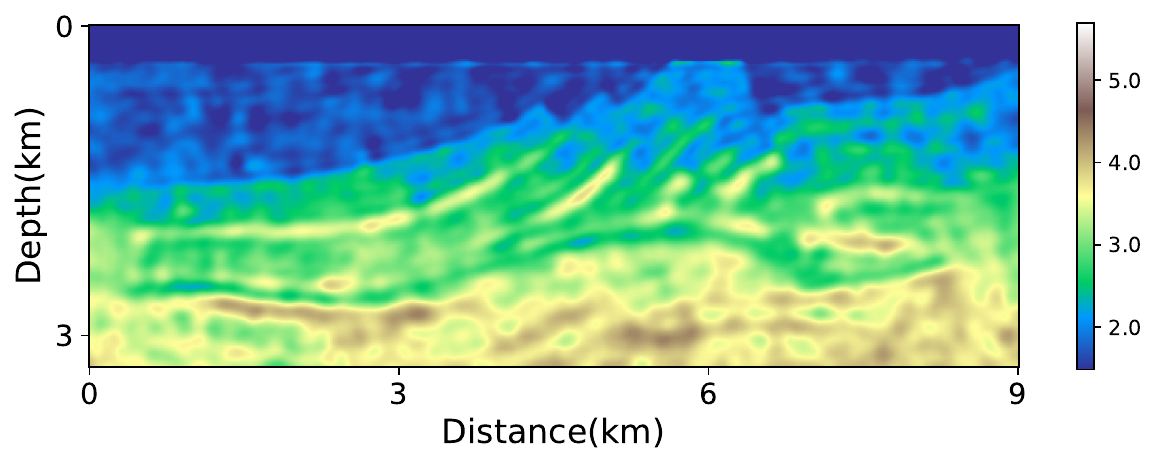}
    \includegraphics[width=0.32\textwidth]{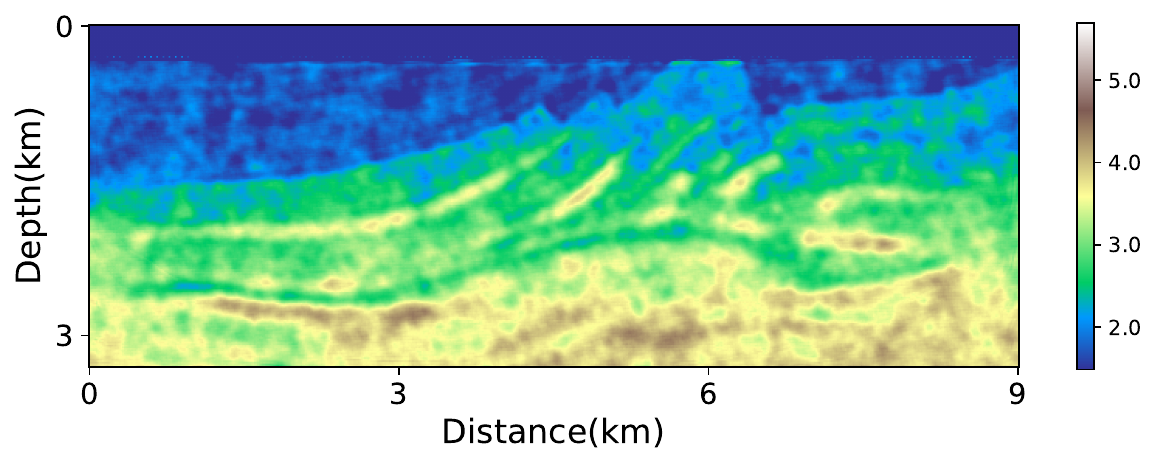}
    \includegraphics[width=0.32\textwidth]{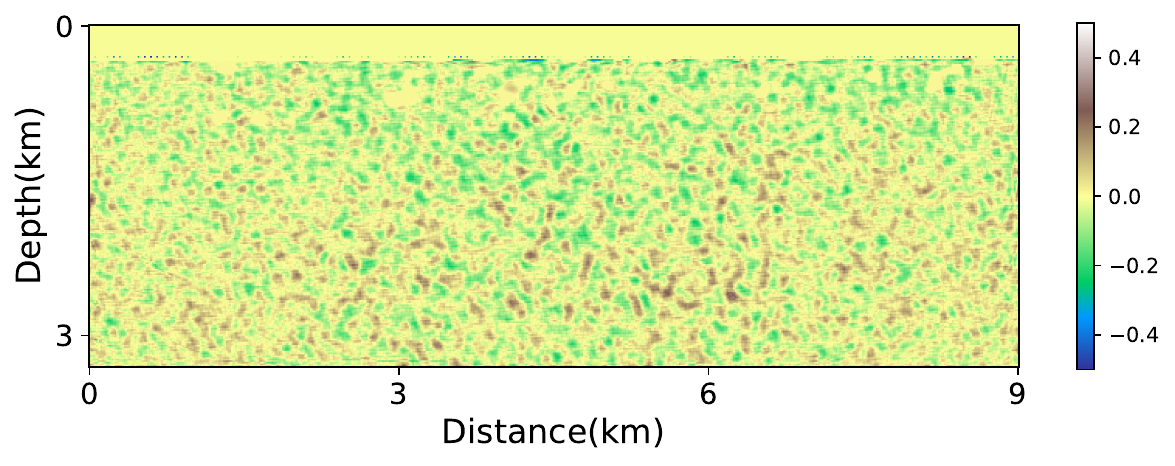}
    \includegraphics[width=0.32\textwidth]{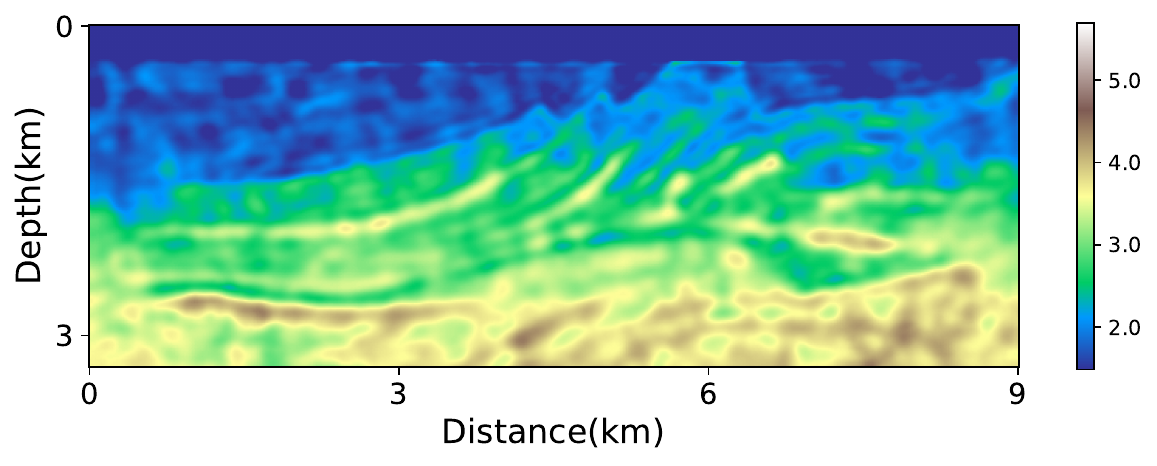}
    \includegraphics[width=0.32\textwidth]{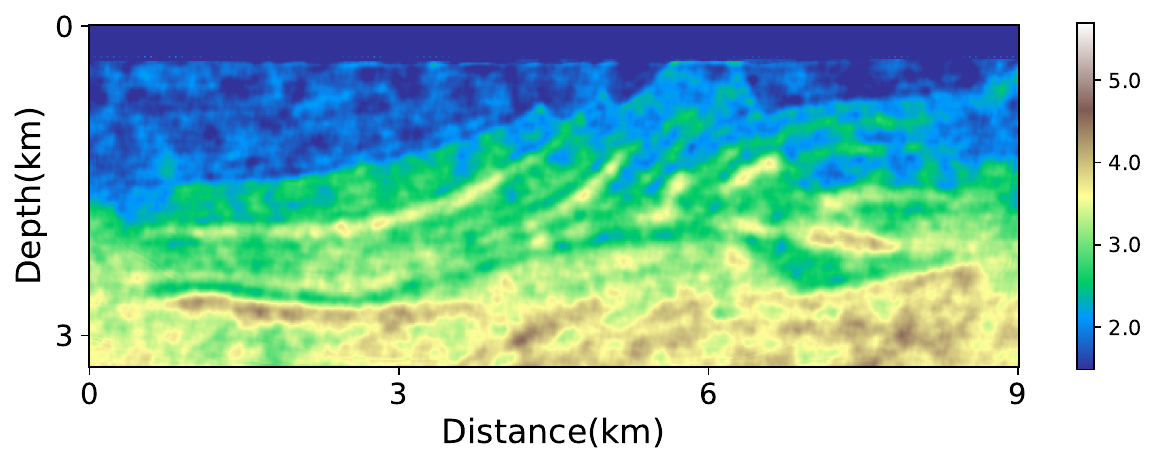}
    \includegraphics[width=0.32\textwidth]{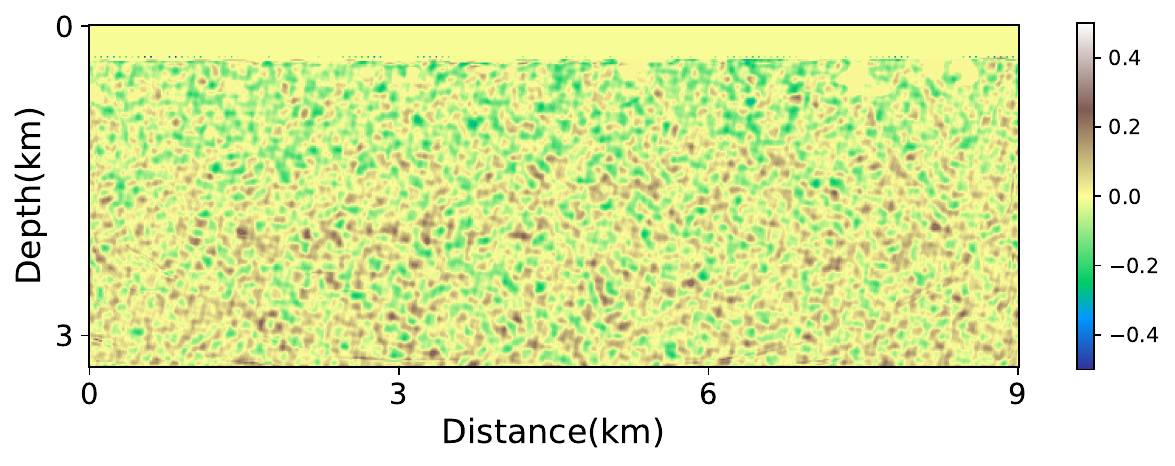}
    \includegraphics[width=0.32\textwidth]{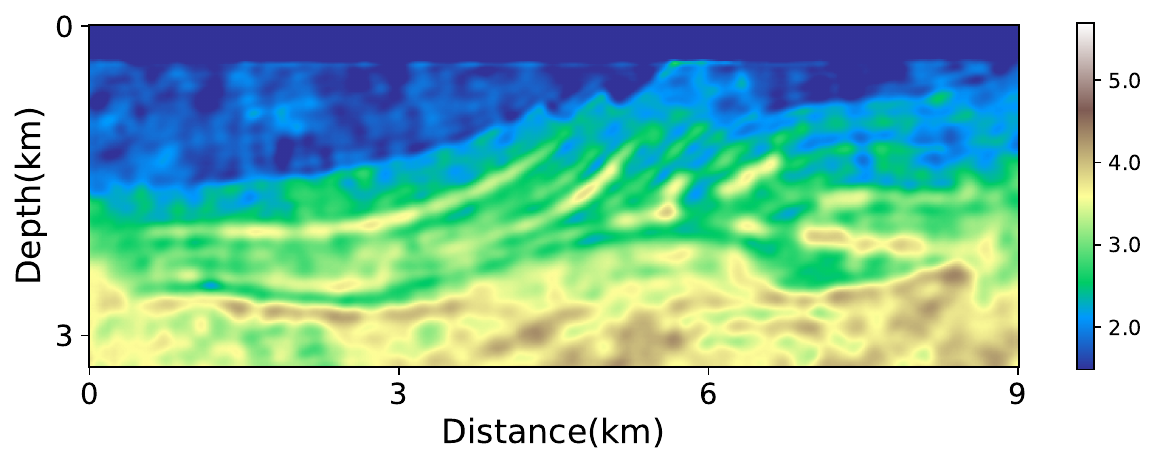}
    \includegraphics[width=0.32\textwidth]{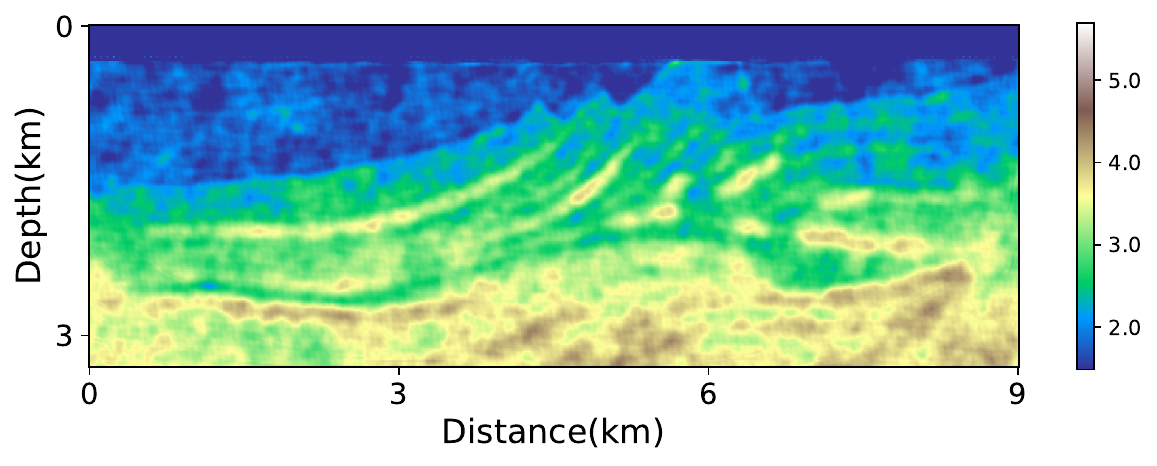}
    \includegraphics[width=0.32\textwidth]{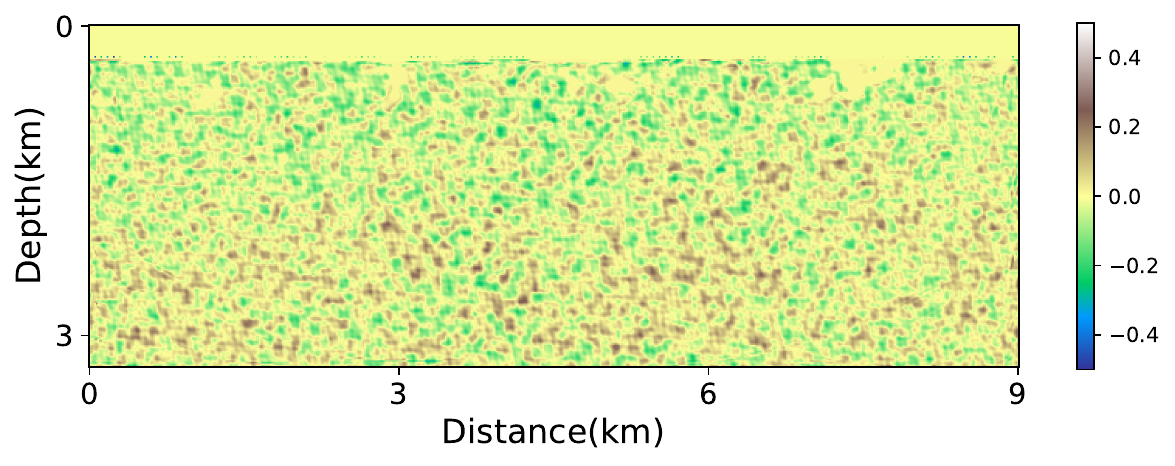}
    \includegraphics[width=0.32\textwidth]{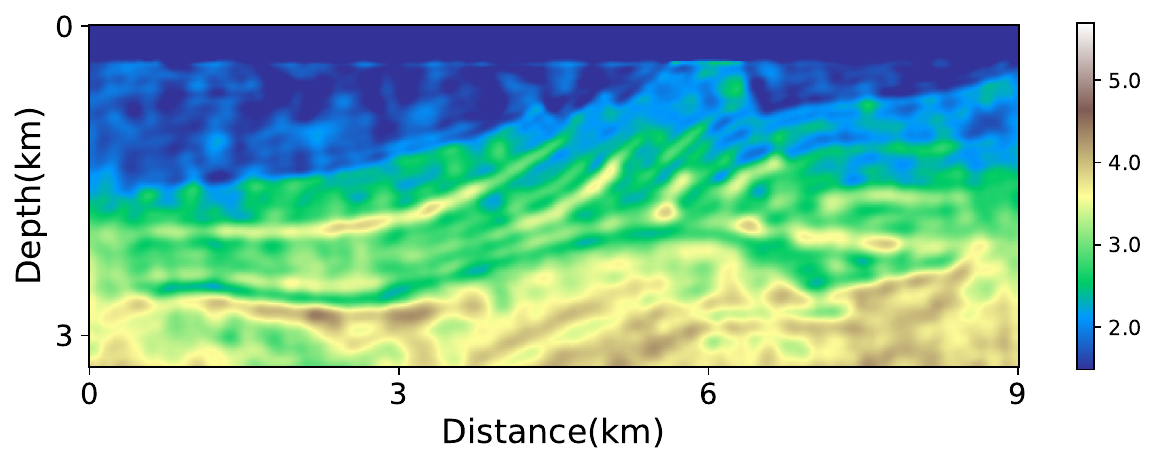}
    \includegraphics[width=0.32\textwidth]{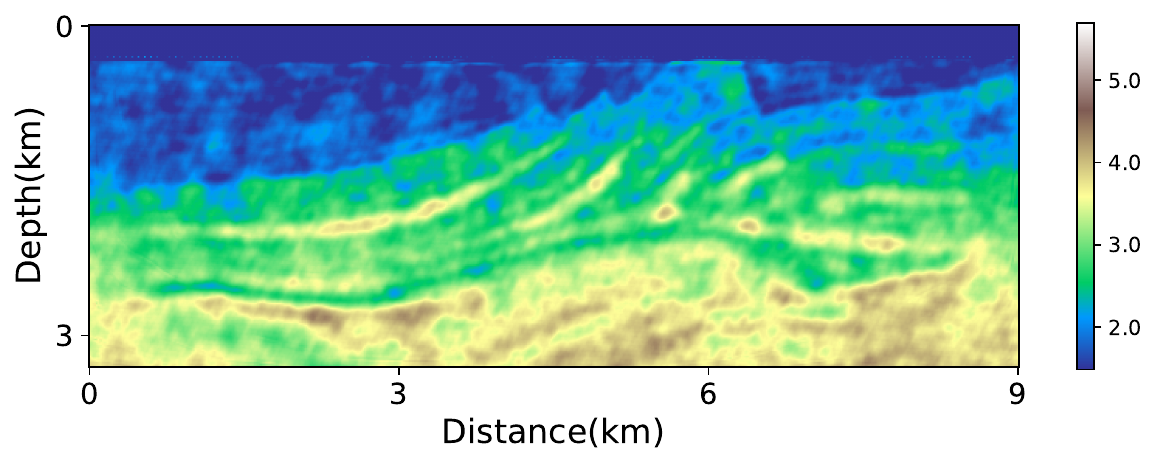}
    \includegraphics[width=0.32\textwidth]{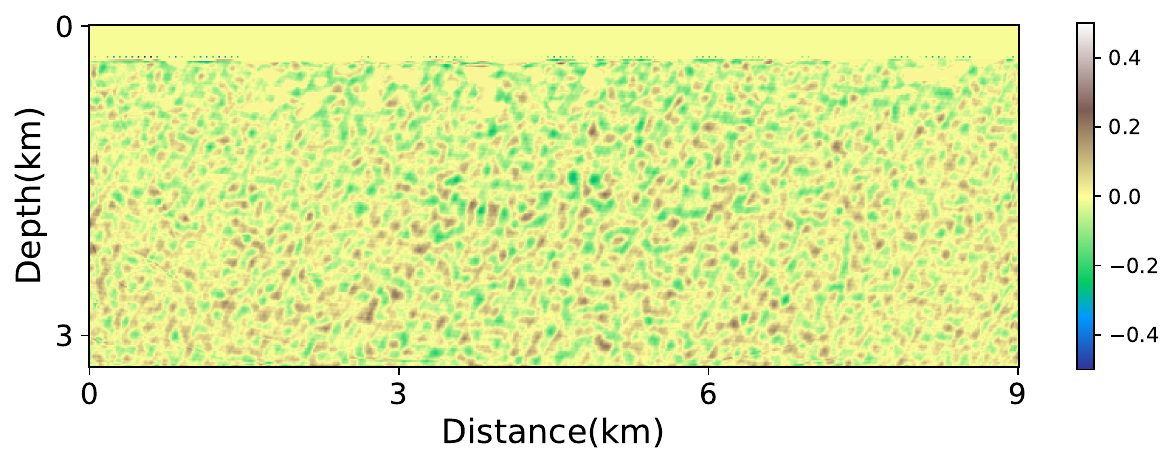}
    \caption{Examples of the generated models from the pre-training process. The first column denotes the generated models with different conditions, the second column represents the target models, and the third column is the error between the target and generated models.}
    \label{fig:3-pretrain-result}
\end{figure*}

%% file: 03_Numerical_Examples.tex
\section{Numerical Examples}
In this section, we evaluate the performance of the proposed conditional CNN on the Marmousi model. The Marmousi model has a size of 221 $\times$ 601 with a grid interval of 15 m in both directions. The source is given by a Ricker wavelet with a peak frequency of 6 Hz. The time step is 0.1 ms and the total recording time is 4 s. There are 15 shots and 300 receivers equally spaced across the model on the surface. We generated the GRF within a range of $\pm$300 m/s with parameter $\lambda=0.1$ and the smoothness coefficient $\eta$ is set to be $1.25$. Since performing FWI for a pollution of model is expensive, we will use the inverted model from a deterministic FWI as our starting point. This will help us reduce the number of iterations and focus on uncertainty quantification \cite{izzatullah2023frugal}. So, we add the GRF perturbations to the deterministic model to produce the initial models for the pretraining stage.

\begin{figure*}[!tb]
\centering
\subfigure{\includegraphics[width=0.3\textwidth]{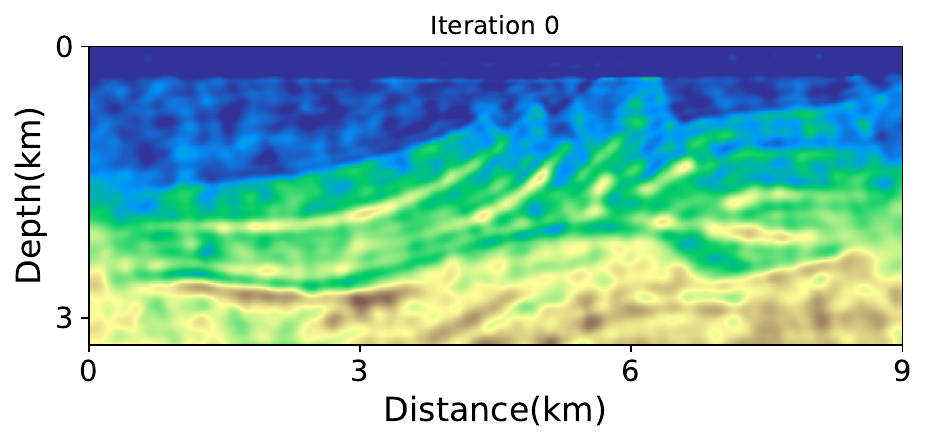}}
\subfigure{\includegraphics[width=0.3\textwidth]{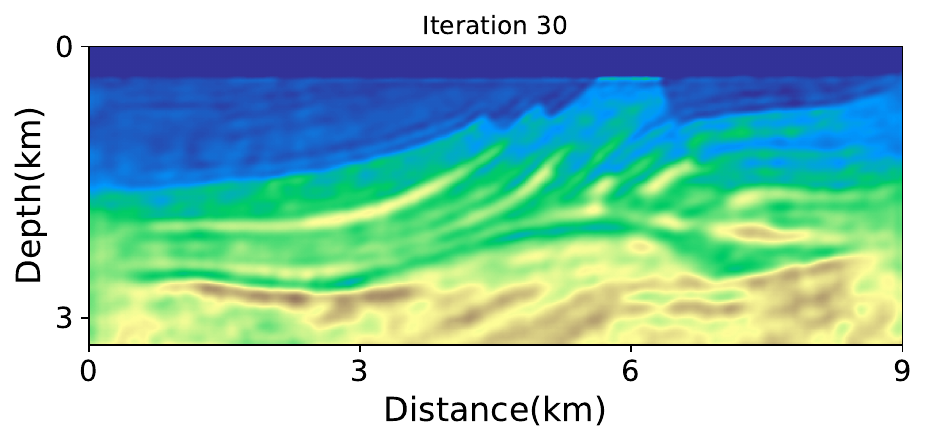}}
\subfigure{\includegraphics[width=0.3\textwidth]{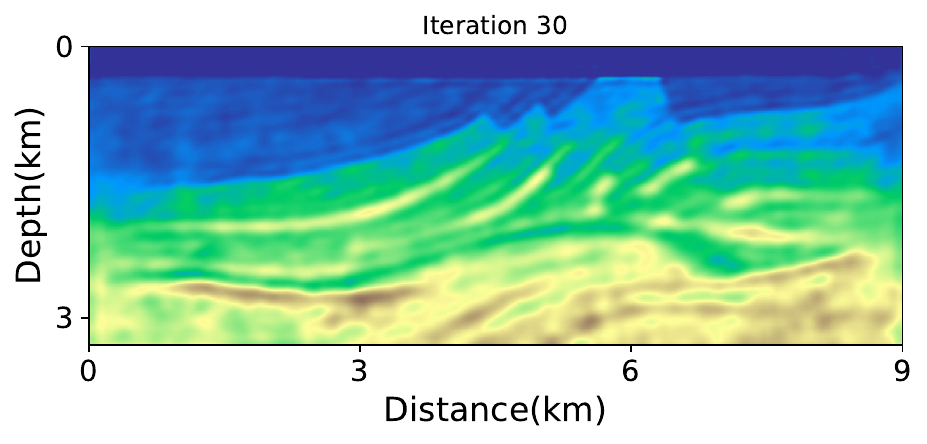}}\\
\subfigure{\includegraphics[width=0.3\textwidth]{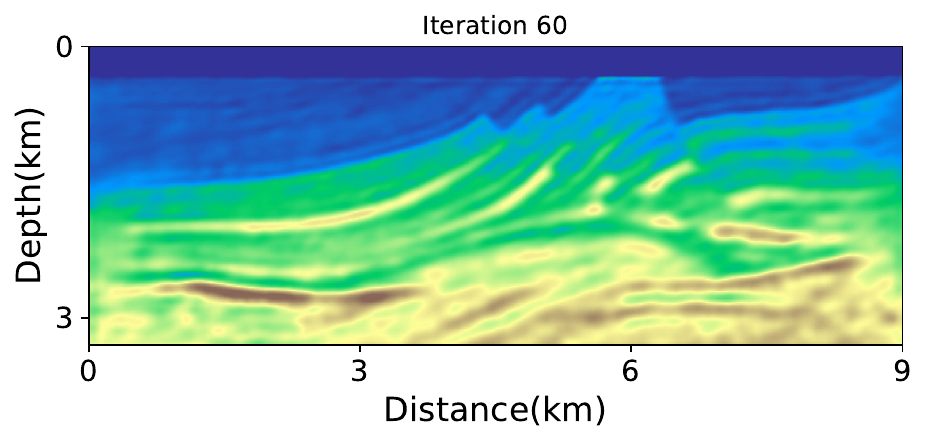}}
\subfigure{\includegraphics[width=0.3\textwidth]{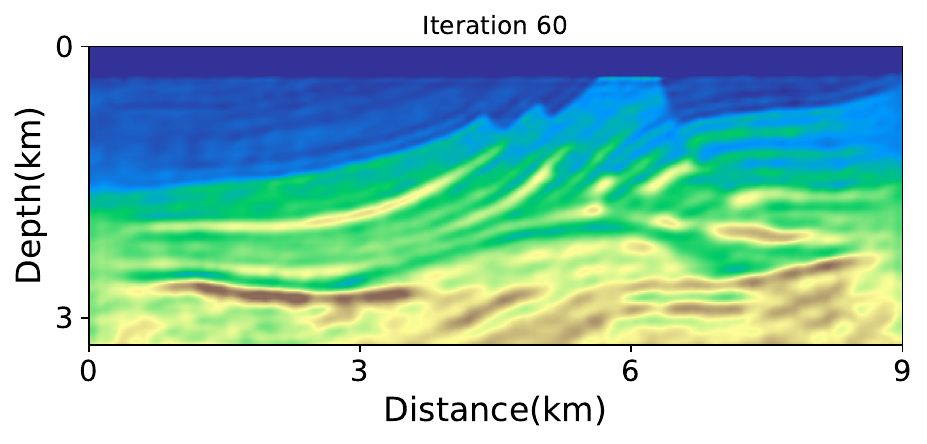}}
\subfigure{\includegraphics[width=0.3\textwidth]{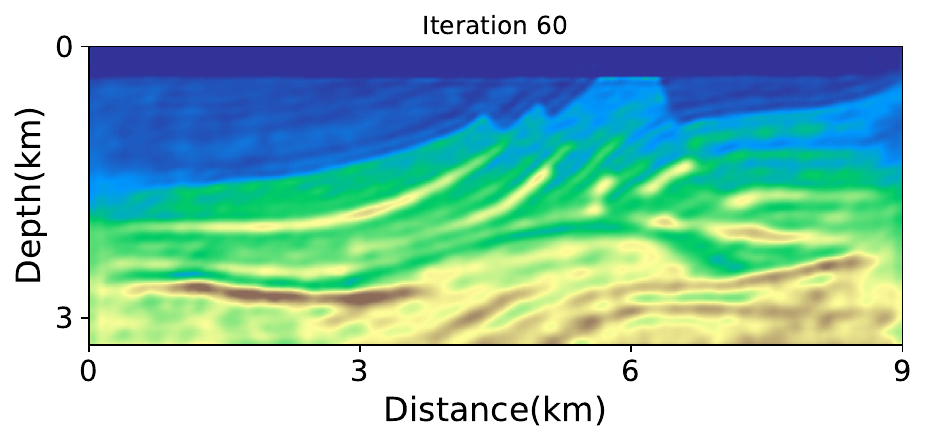}}\\
\subfigure{\includegraphics[width=0.3\textwidth]{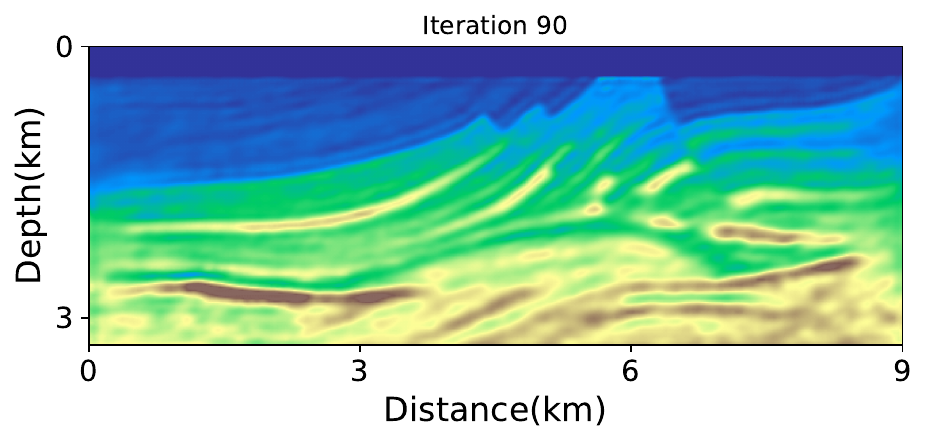}}
\subfigure{\includegraphics[width=0.3\textwidth]{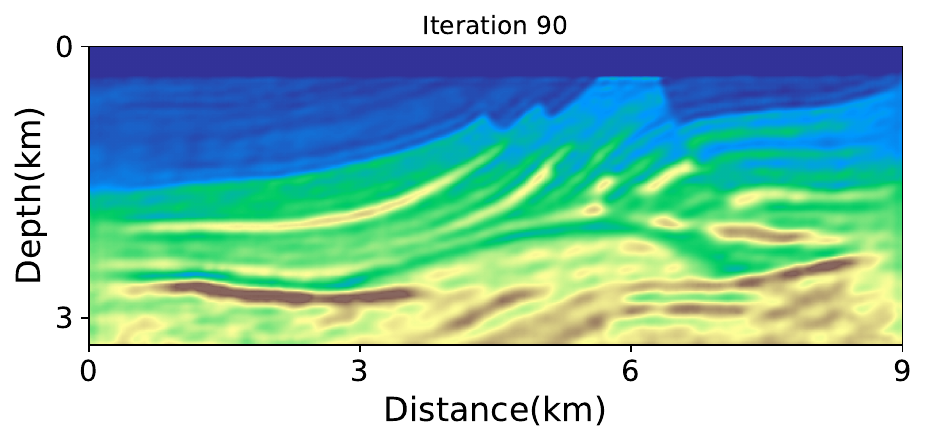}}
\subfigure{\includegraphics[width=0.3\textwidth]{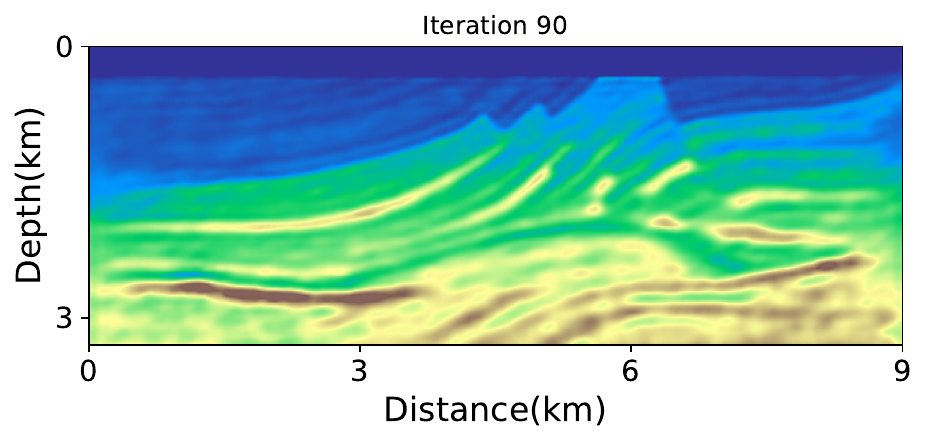}}\\
\subfigure{\includegraphics[width=0.3\textwidth]{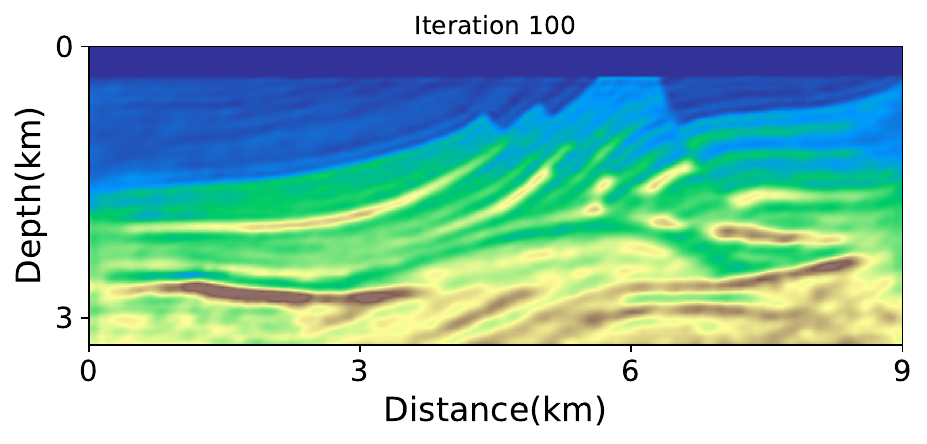}}
\subfigure{\includegraphics[width=0.3\textwidth]{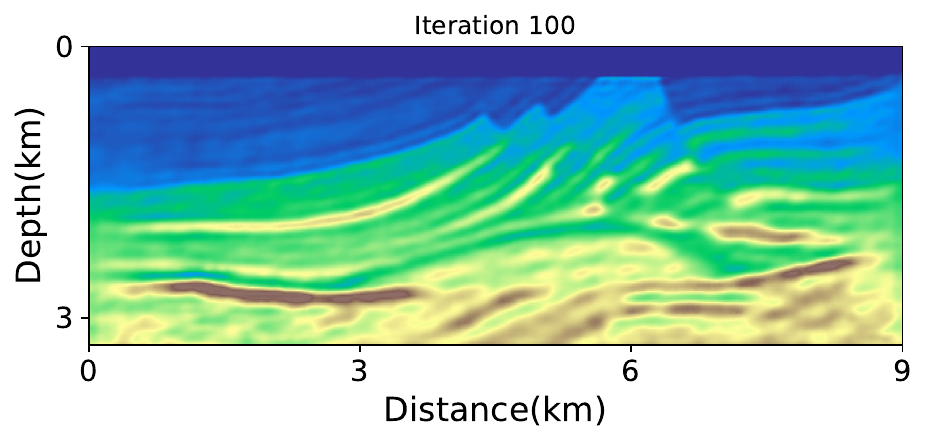}}
\subfigure{\includegraphics[width=0.3\textwidth]{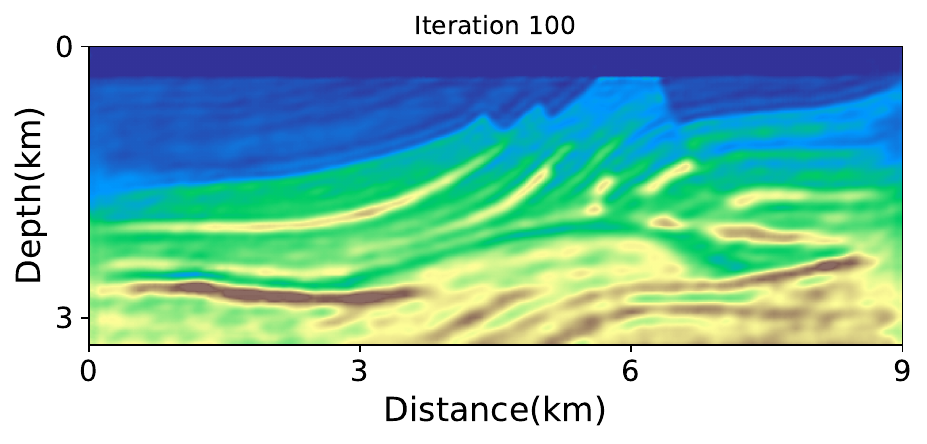}}\\
\subfigure{\includegraphics[width=0.4\textwidth]{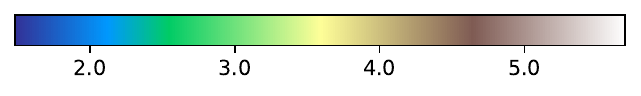}}
\caption{Examples of the inverted models from the FWI stage. The first column represents the inverted models at different iterations under condition 0, the second column shows the inverted models at different iterations under condition 20, and the third column denotes the inverted models at different iterations under condition 40.}
\label{fig:5-condition-FWI-result}
\end{figure*}

\subsection{The Pretraining Stage}
The proposed CNN architecture follows a U-Net design, comprising five layers in each of the encoder and decoder parts, with a total of 0.5 million parameters. In the pretraining stage (Figure \ref{fig:2-workflow-of-cnn}), the Adam optimizer is employed to optimize the network’s parameters, while the $L_2$ loss function is utilized to quantify the misfit between the network’s output and the target model across different conditions. The input to the network comprises two components: random noise and a vector carrying the condition information. Specifically, as described earlier, a one-hot vector, with a length equal to the number of samples/particles representing the prior, is used as the condition input. For improved conditional results, we actually integrate the one-hot vector into each of the U-net layers, ensuring that the neural network effectively learns target models. Figure \ref{fig:3-pretrain-result} shows examples of the predicted models, target models, and the difference between them for a number of conditions. Notably, the difference between the predicted model and the target model is negligible. From this result, we notice that the conditional CNN has successfully learned different target models under varying conditions.
\begin{figure*}[!tb]
    \centering
    \includegraphics[width=\textwidth]{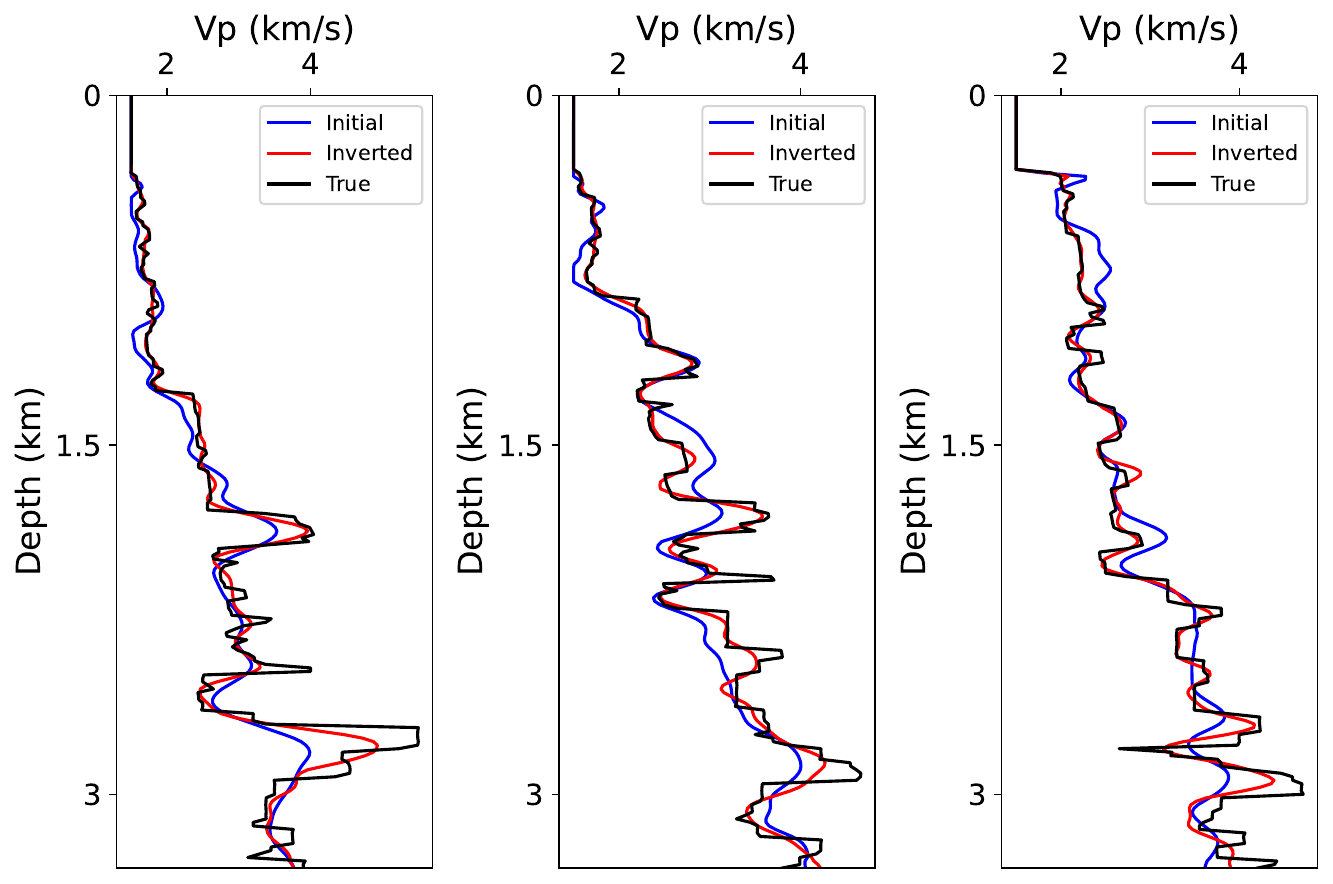}
    \caption{Velocity profile at distances of 3, 4.5, and 6 km of condition 20 from left to right, respectively. The solid black, blue, and red lines represent the true, initial, and inverted velocities models, respectively.}
    \label{fig:6-lt}
\end{figure*}

The elements of the condition vector act as a latent representation of the GRF features. In order to demonstrate this feature of conditional CNN, we test alternative values for the 50-elements condition vector we use. So, we have trained with the network using 50 velocity model samples from the prior, assigning a unique one-hot vector for each sample. Now, we use alternative vectors in which the sum of the elements are equal to one to maintain the distribution property of the condition. As shown in Figure \ref{fig:4-alternative-vector} (a) and (d) corresponding to condition vectors [0.8, 0.2, ..., 0] and [0.2, 0.8, ..., 0] respectively, the velocity models generated by the proposed conditional CNN look similar to the other prior models. Figure \ref{fig:4-alternative-vector} (b, e) and (c, d) show the difference between these newly generated velocity models and the trained ones corresponding to one-hot vector [1, 0,..., 0] (second column) and [0, 1, 0,..., 0] (third column). We note, as would be expected, that the difference is smaller when their condition vectors are closer to each other. In addition, there are no obvious structural features in the difference, which implies that the conditional CNN has captured the main features of the underlying model (without GRF). This diversity observed in the generated velocity models presents avenues for implementing sampling strategies to extract more comprehensive insights from the network, thereby enhancing the results of the posterior distribution.\\
\begin{figure*}[!tb]
    \centering
    \includegraphics[width=0.48\textwidth]{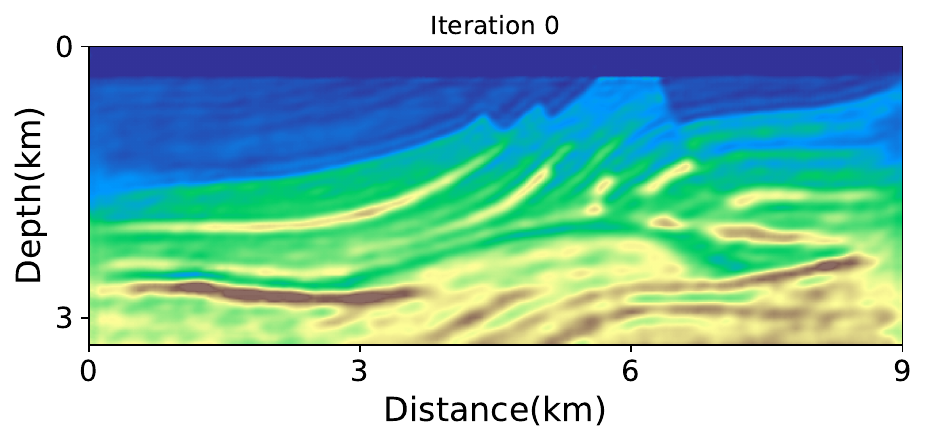}
    \includegraphics[width=0.48\textwidth]{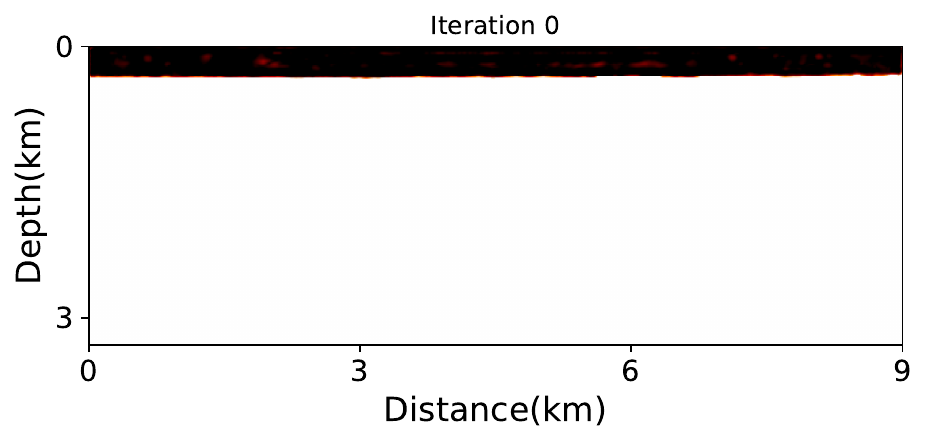}
    \includegraphics[width=0.48\textwidth]{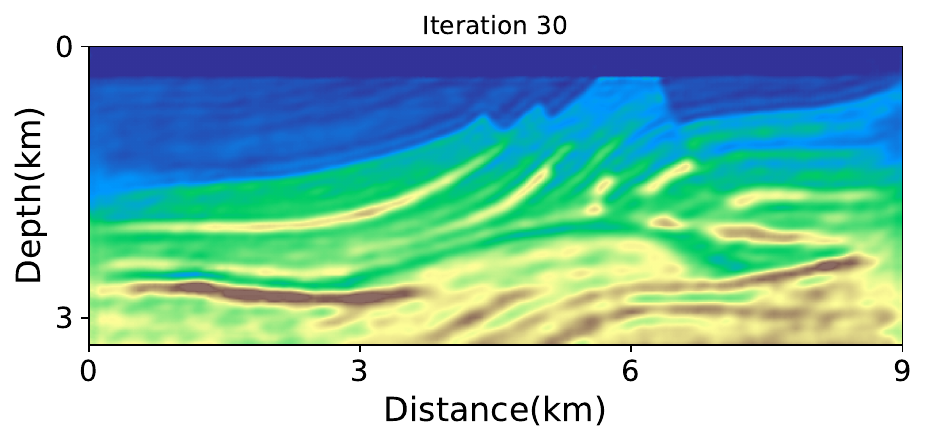}
    \includegraphics[width=0.48\textwidth]{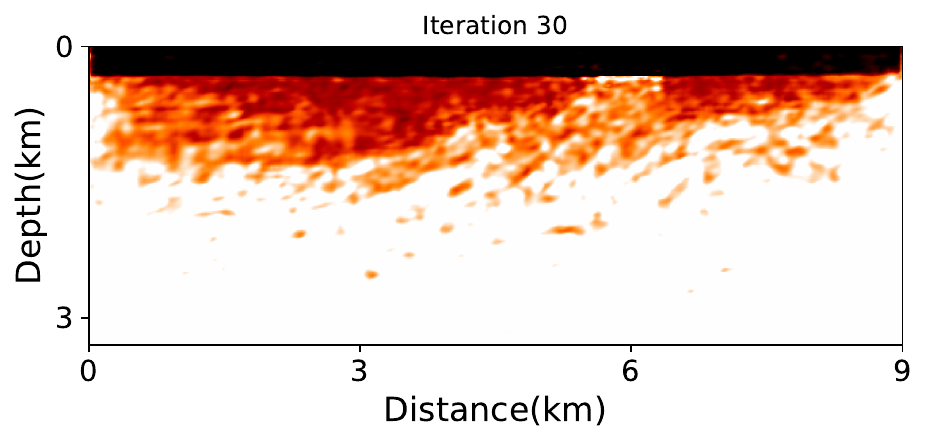}
    \includegraphics[width=0.48\textwidth]{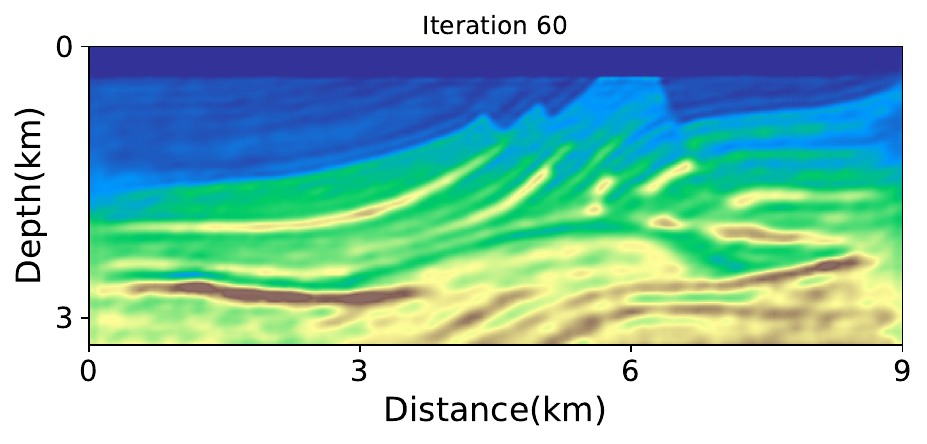}
    \includegraphics[width=0.48\textwidth]{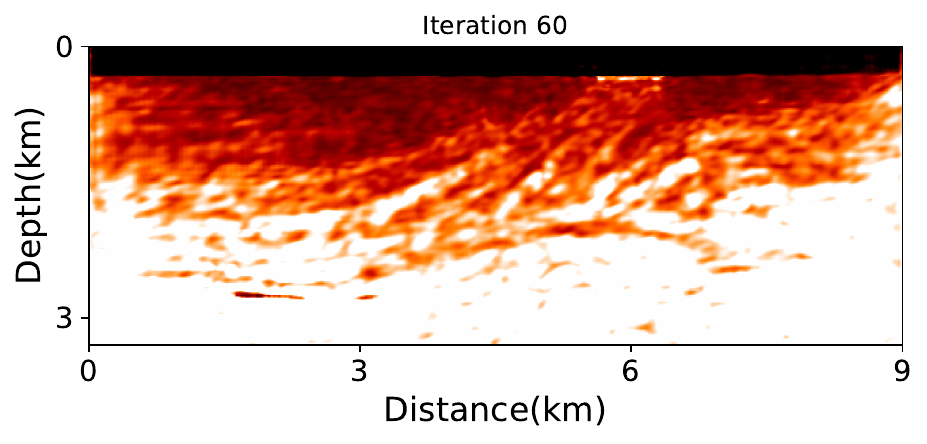}
    \includegraphics[width=0.48\textwidth]{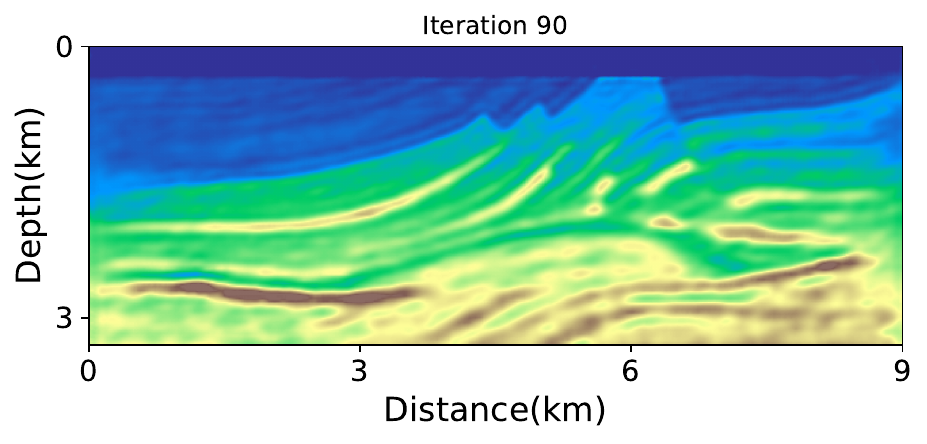}
    \includegraphics[width=0.48\textwidth]{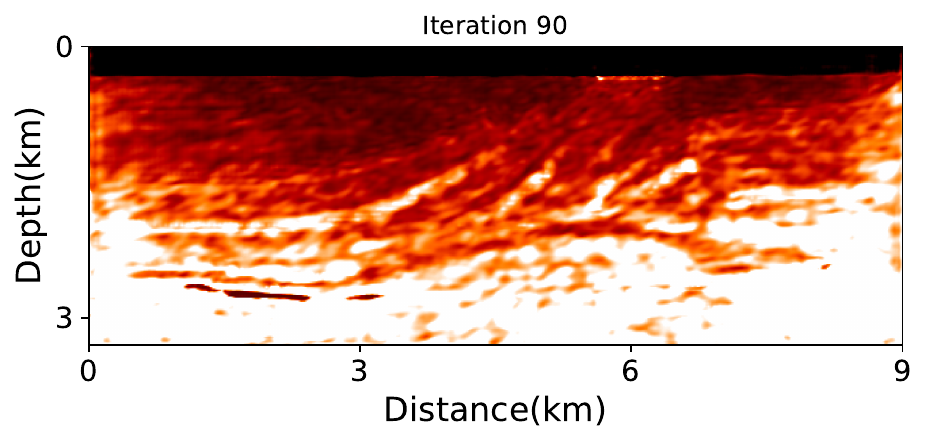}
    \includegraphics[width=0.48\textwidth]{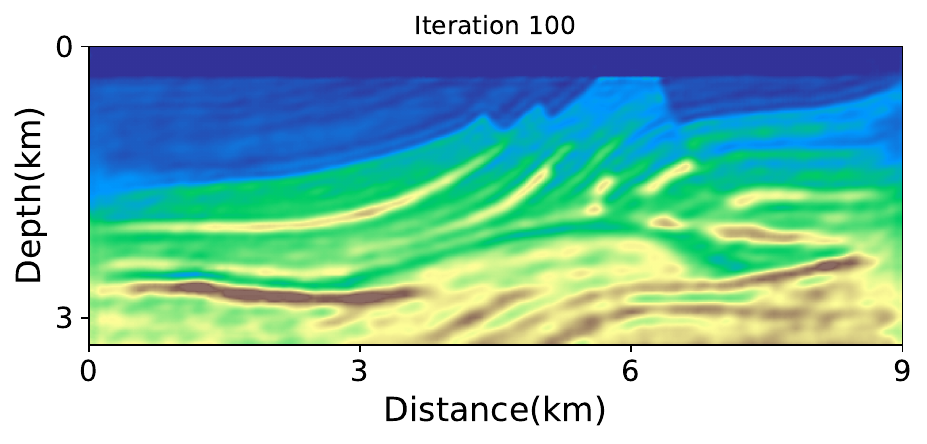}
    \includegraphics[width=0.48\textwidth]{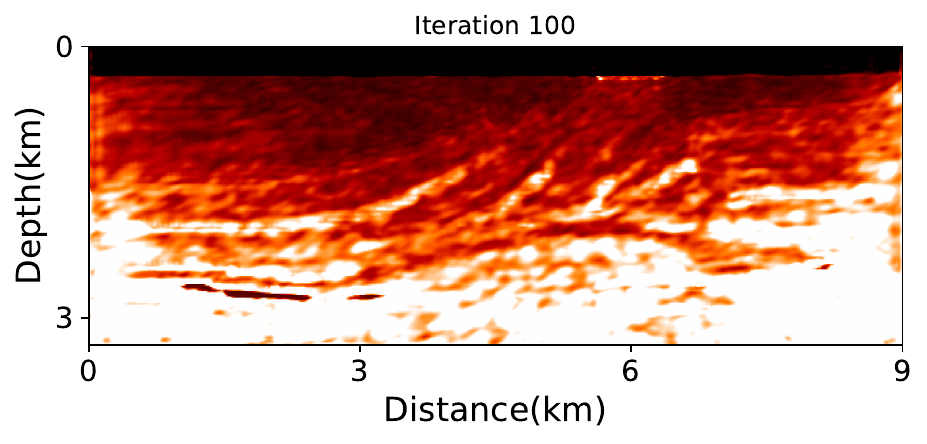}
    \includegraphics[width=0.4\textwidth]{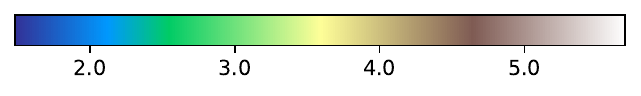}
    \includegraphics[width=0.4\textwidth]{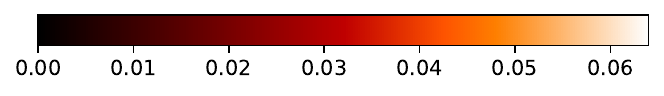}     
    \caption{Examples of the mean and standard deviation for the inverted models from the FWI stage at different iterations. }
    \label{fig:7-lt}
\end{figure*}

\subsection{The FWI Stage}
After the pretraining, we conducted 100 FWI iterations to update the conditional CNN model representation of the priors so that it could generate samples representing the posterior distribution. At the final iteration, we aggregated all inverted models and computed their mean and standard deviation. The inverted velocities are shown in Figures \ref{fig:5-condition-FWI-result} under different iterations, and we can observe that with the increase of iteration, the GRF can be effectively removed. Moreover, and for all the conditions the velocity models converge close to the mean velocity. Figure \ref{fig:6-lt} shows a comparison between the traces extracted from figure \ref{fig:5-condition-FWI-result} at a horizontal distance of 3, 4.5, and 6 km for condition 20 (implying the one-hot vector has a value of one only at element 20). In Figure \ref{fig:6-lt}, the solid black, blue, and red lines represent the true, initial, and inverted velocities, respectively. We can see that there is a good recovery on the scattering part of the velocity model. These spectra further demonstrate that the proposed conditional CNN FWI can effectively remove the GRF perturbation and converge to the true model.

Figure \ref{fig:7-lt} shows the mean and standard deviation of the posterior obtained through the application of the proposed conditional CNN network. As expected, we obtain higher standard deviation in areas where the illumination is probably weak, specifically the deeper parts and along the edges due to limitations in acquisition. We also note that we obtain higher uncertainty in regions of higher velocity as the resolution decreases with increasing wavelength. Finally, it is noteworthy to mention that there were minimal difference between the standard derivation of iterations 90 and 100, suggesting convergence of the conditional CNN network.
\\
\begin{figure*}[!tb]
    \centering
    \includegraphics[width=0.42\textwidth]{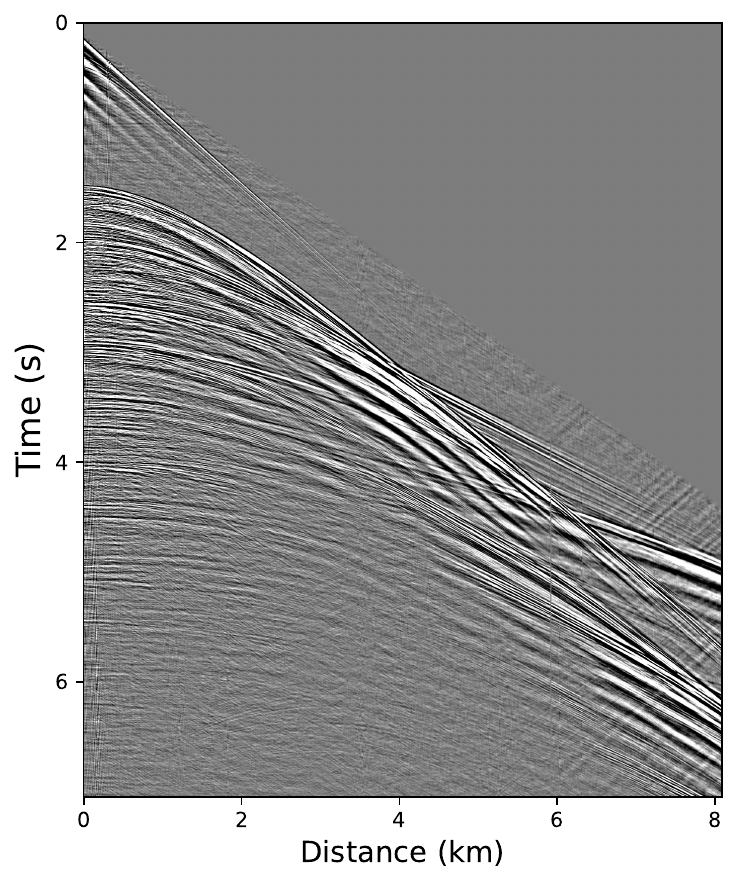}
    \includegraphics[width=0.42\textwidth]{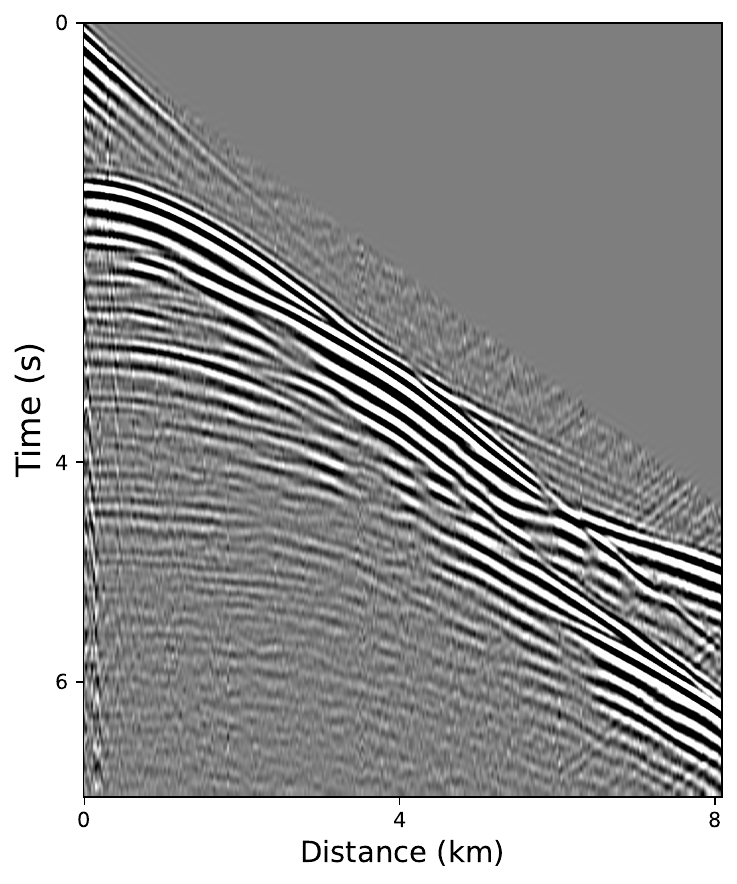}
    \caption{Raw and filtered shot gathers of filed data shown on the left and right, respectively.}
    \label{fig:9-lt}
\end{figure*}

\begin{figure*}[!tb]
    \centering
    \includegraphics[width=0.32\textwidth]{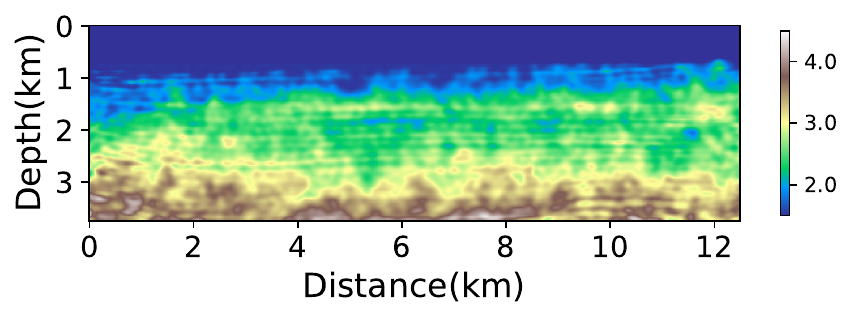}
    \includegraphics[width=0.32\textwidth]{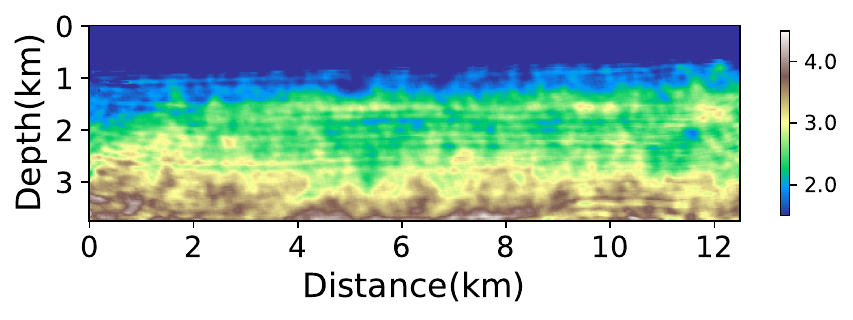}
    \includegraphics[width=0.32\textwidth]{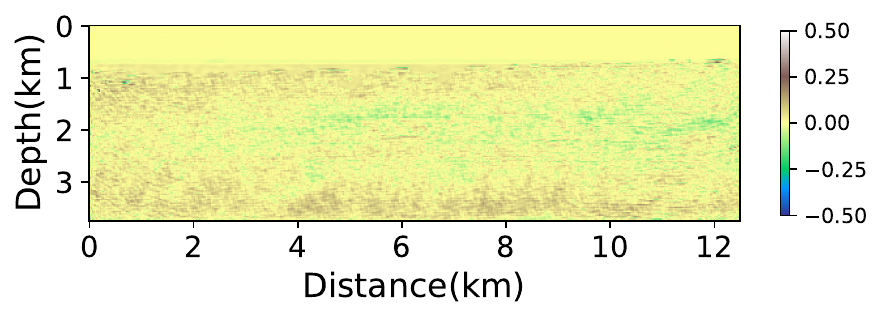}
    \includegraphics[width=0.32\textwidth]{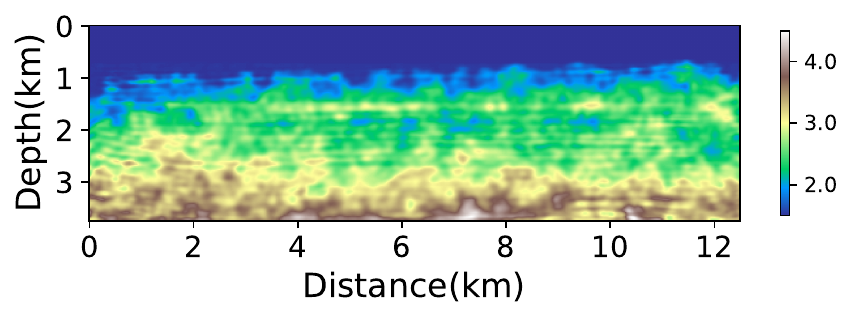}
    \includegraphics[width=0.32\textwidth]{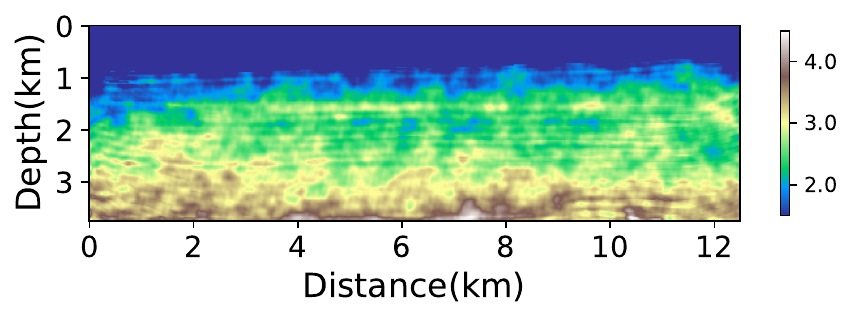}
    \includegraphics[width=0.32\textwidth]{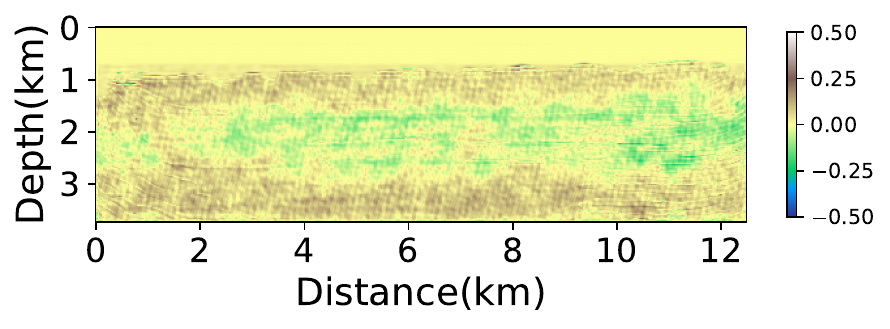}    
    \includegraphics[width=0.32\textwidth]{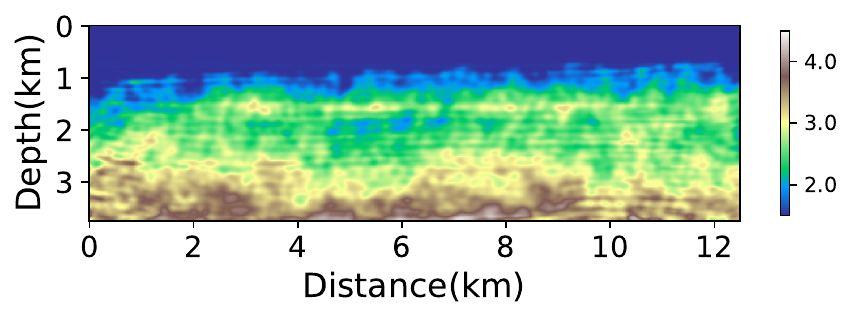}
    \includegraphics[width=0.32\textwidth]{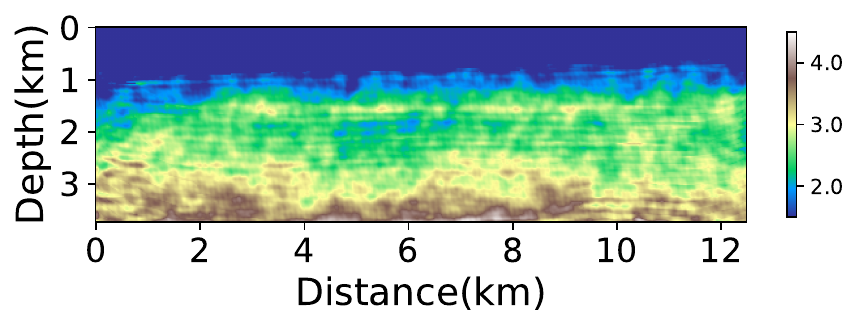}
    \includegraphics[width=0.32\textwidth]{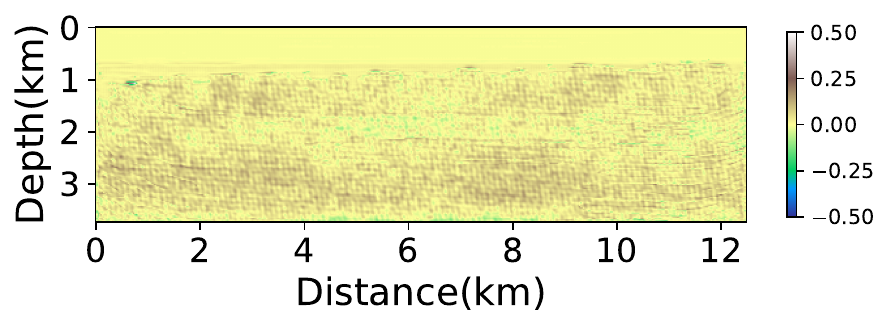}
    \caption{Examples of the generated models from the pre-training process. The first column denotes the generated models with different conditions, the second column represents the target models, and the third column is the difference between the target and generated models.}
    \label{fig:9-condition-FWI-result}
\end{figure*}

%% file: 04_Field_Data_Application.tex
\section{Field Data Application}
Here, we use a 2-D field marine dataset from the North-Western Australia Continental shelf acquired by CGG to validate our method. We choose 116 shot gathers among the original 1824 shot gathers with a 18.75 m shot interval. Each shot gather includes 324 receivers at a 25 m spacing interval. The maximum recording time is about 7 s with 1 ms sampling interval. As shown in Figure \ref{fig:9-lt}, we first applied a low pass filter on the field data with a cut-off frequency of 10.5 Hz.

Here, we use the inversion result obtained by \cite{saad2024siamesefwi}, as our starting model. The velocity model is designed to be 12.5 km long and 3 km deep, both with 25 m grid interval. We use the same conditional CNN network parameters as those employed in the previous Marmousi model example. Figure \ref{fig:9-condition-FWI-result} shows examples of the predicted and target models with different conditions. The difference between the predicted and the target models is negligible. After the pretraining, we conducted 100 FWI iterations to update the conditional CNN model representation of the priors, generating samples approximately representing a posterior distribution. 

\begin{figure*}[!tb]
    \centering
    \includegraphics[width=0.32\textwidth]{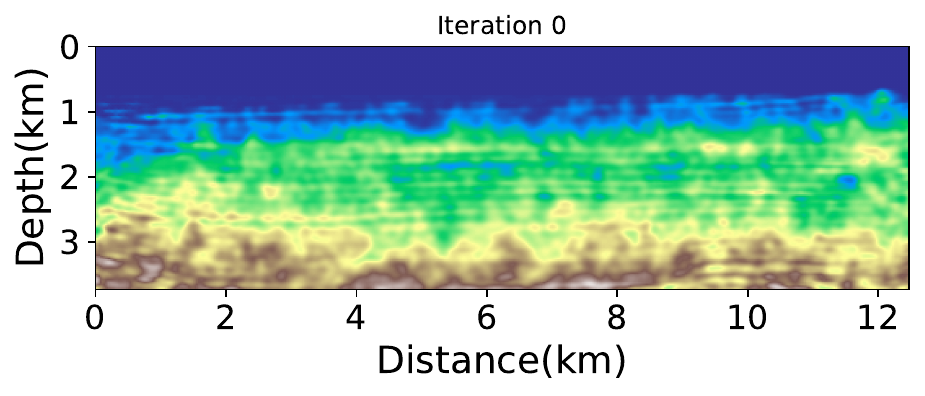}
    \includegraphics[width=0.32\textwidth]{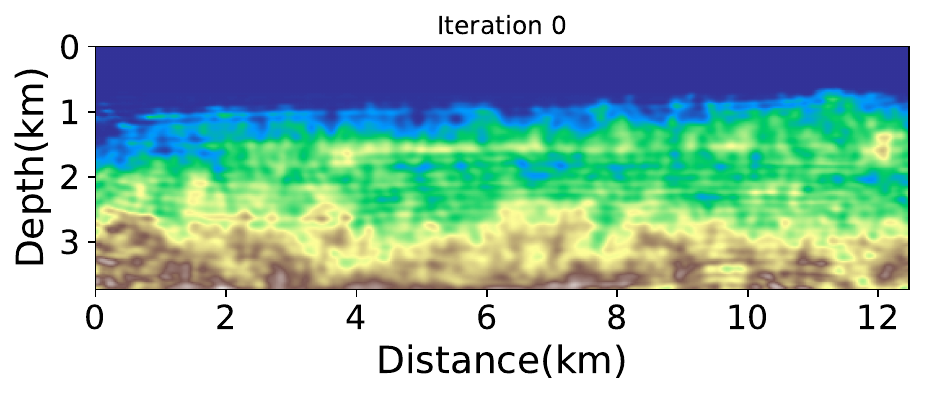}
    \includegraphics[width=0.32\textwidth]{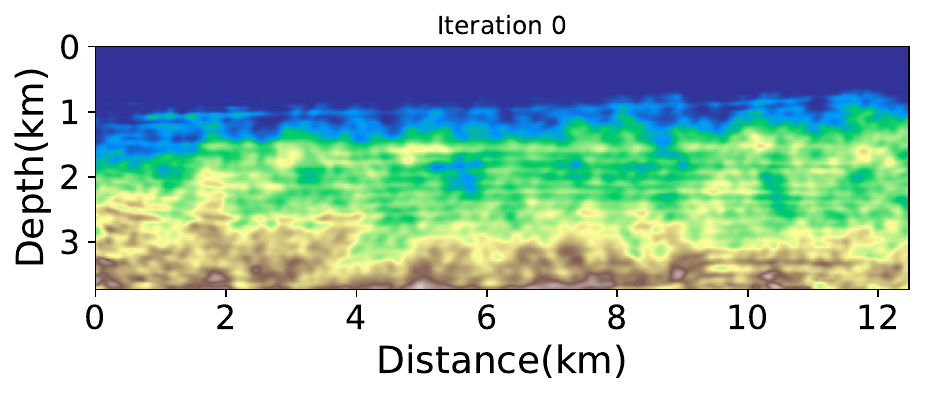}
    \includegraphics[width=0.32\textwidth]{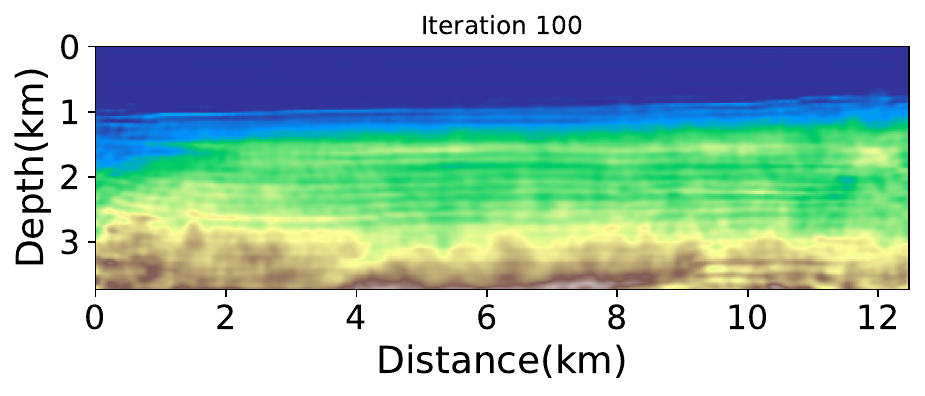}
    \includegraphics[width=0.32\textwidth]{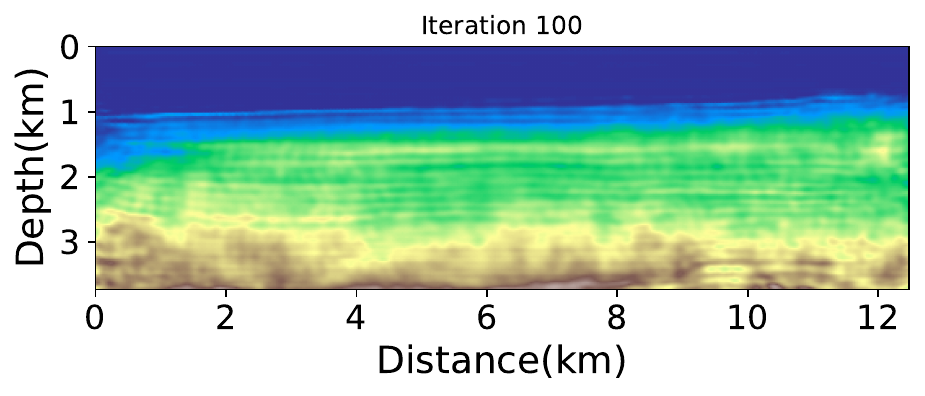}
    \includegraphics[width=0.32\textwidth]{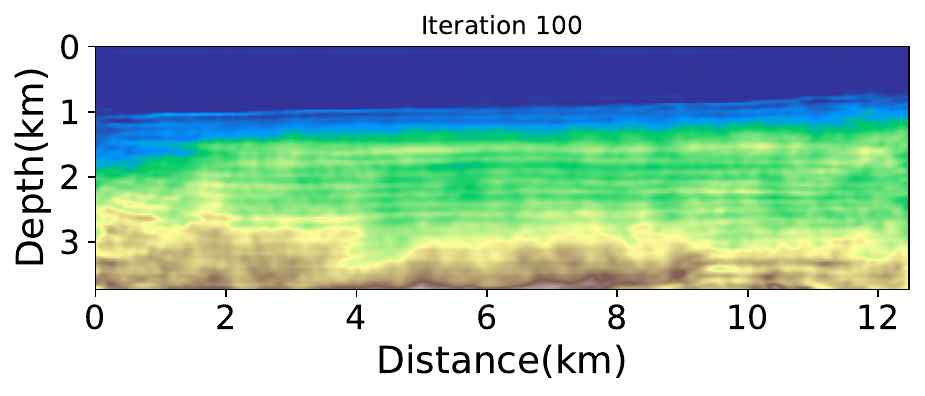}
    \includegraphics[width=0.4\textwidth]{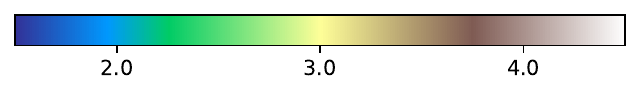}
    \caption{Examples of the inverted models from the FWI stage. The first column represents the inverted models for zero and 100 iterations for condition 0, the second column shows the same for condition 20, and the third column shows the same for condition 40.}
    \label{fig:10-condition-FWI-result}
\end{figure*}

\begin{figure*}[!tb]
    \centering
    \includegraphics[width=0.9\textwidth]{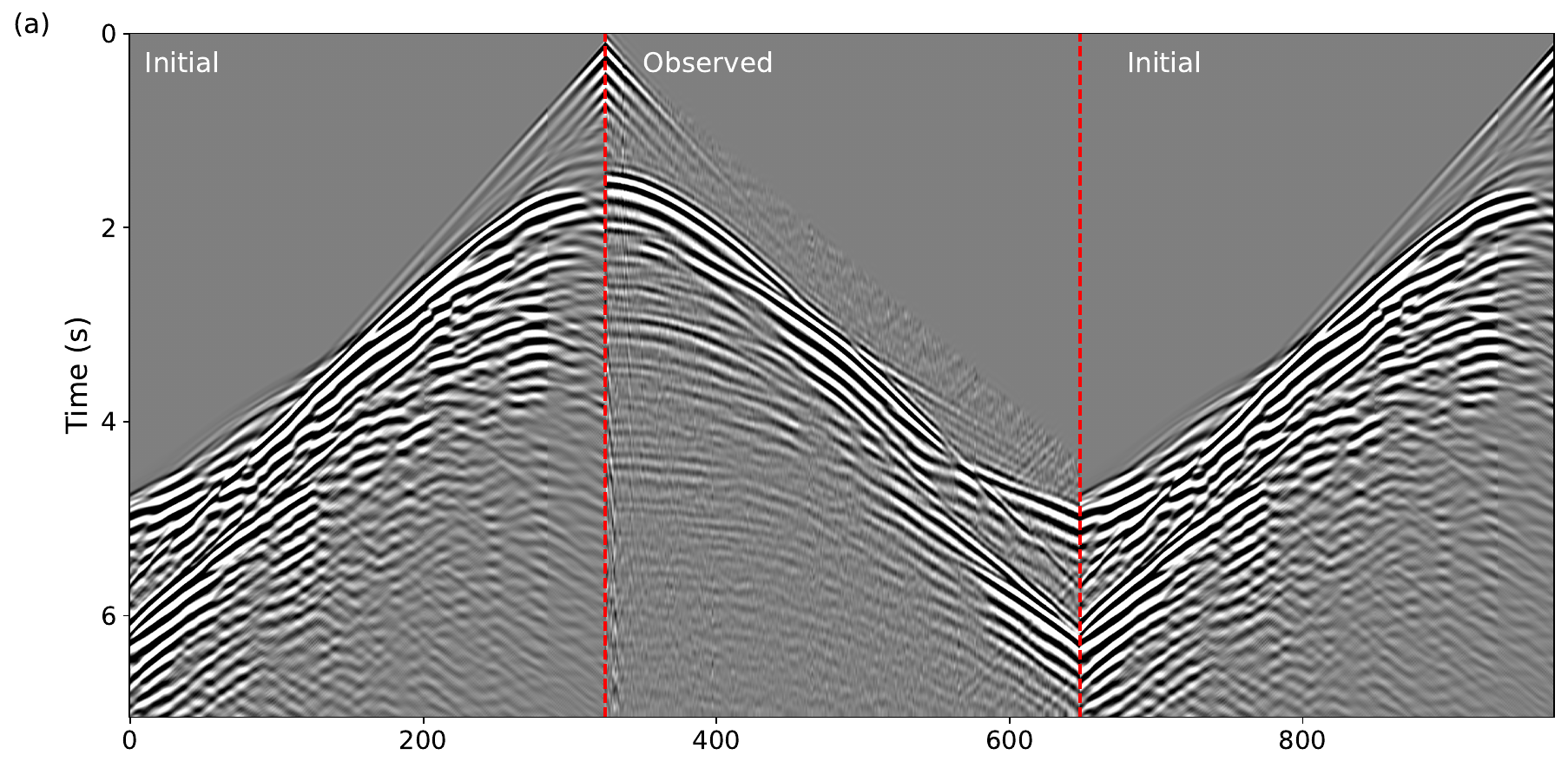}
    \includegraphics[width=0.9\textwidth]{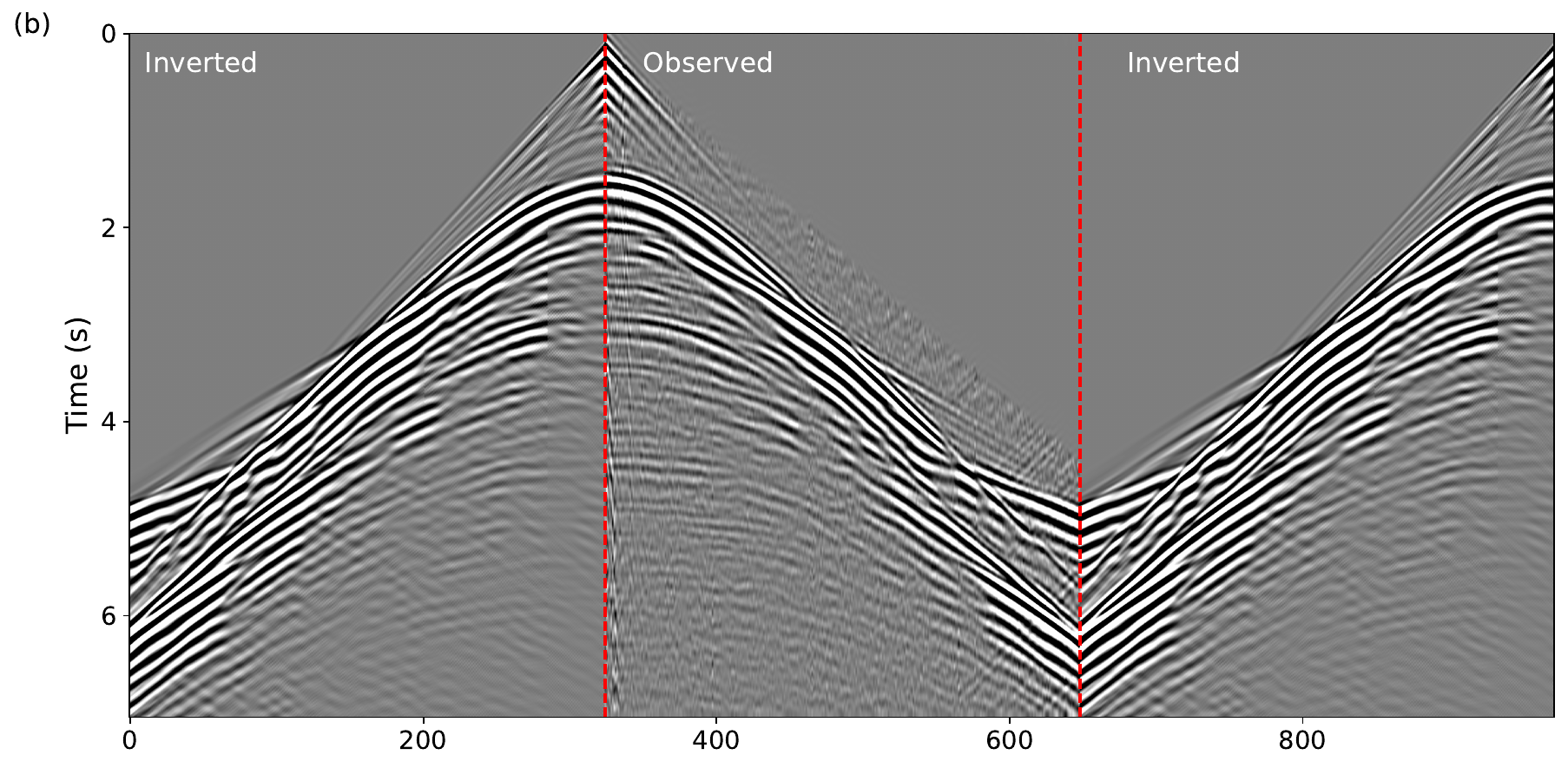}
    \caption{A comparison between common shot gathers for observed data and the modeled data using the (a) initial model, and (b) inverted model, respectively.}
    \label{fig:13-lt}
\end{figure*}

The inverted velocities are shown in Figures \ref{fig:10-condition-FWI-result} across different conditions. We observe that, under all the conditions, the velocity models converge towards the mean velocity. Figure \ref{fig:13-lt} shows the observed shot gather alongside the shot gathers generated from both the initial and inverted model for condition 40. We can see that the predicted and measured data resemble each other well. The alignment of events at both the near and far offsets between the predicted and observed gathers verifies the high accuracy of the inverted model.

\begin{figure*}[!tb]
    \centering
    \includegraphics[width=0.48\textwidth]{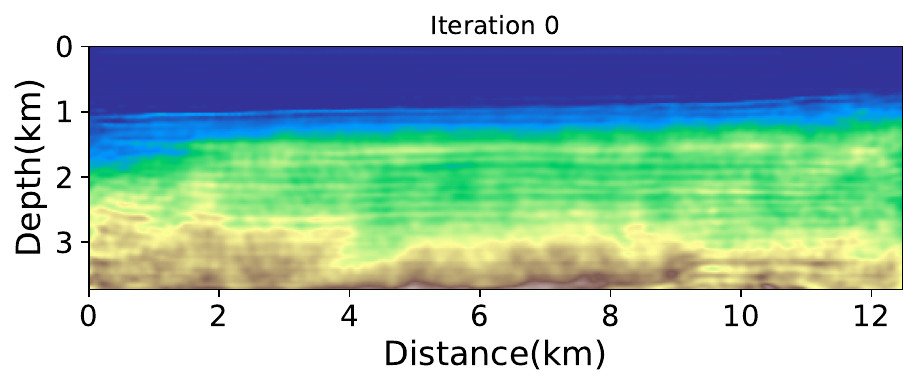}
    \includegraphics[width=0.48\textwidth]{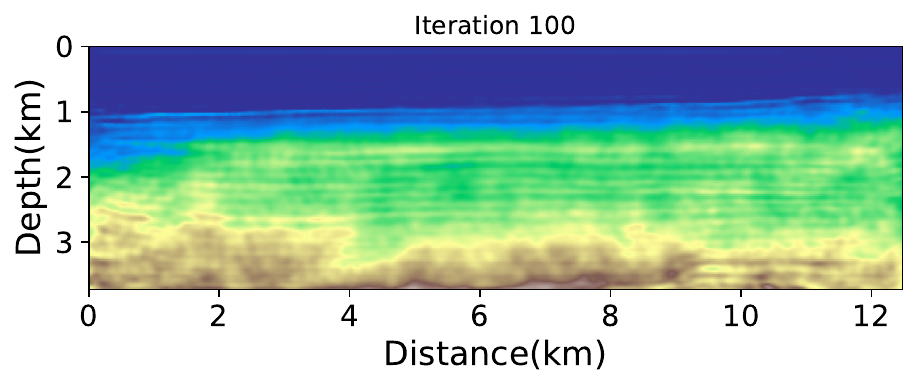}
    \includegraphics[width=0.48\textwidth]{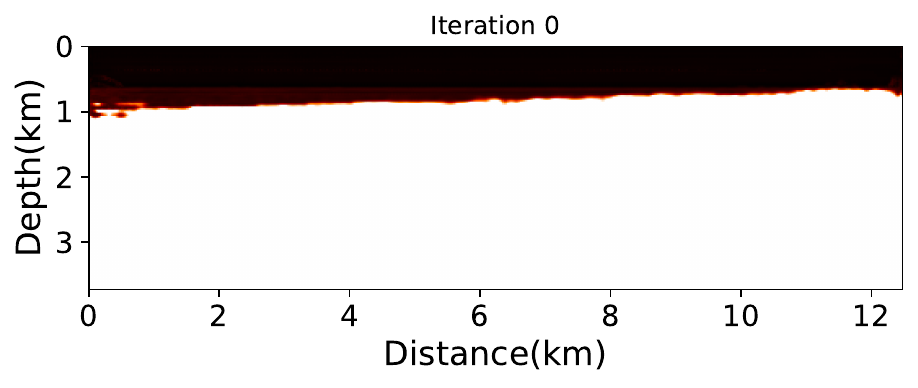}
    \includegraphics[width=0.48\textwidth]{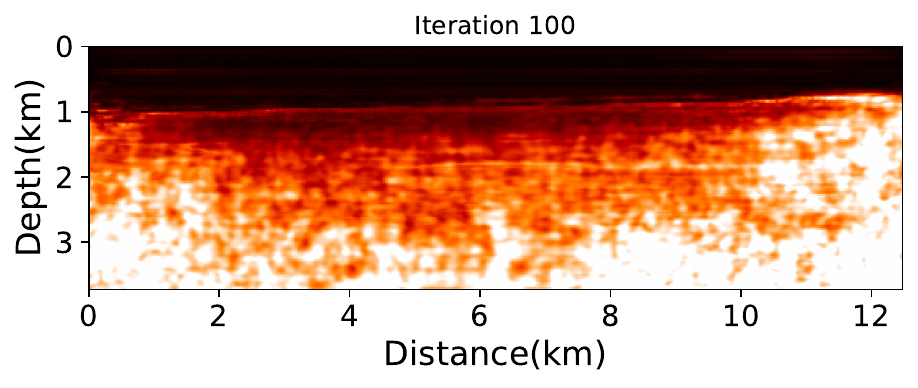}
    \includegraphics[width=0.48\textwidth]{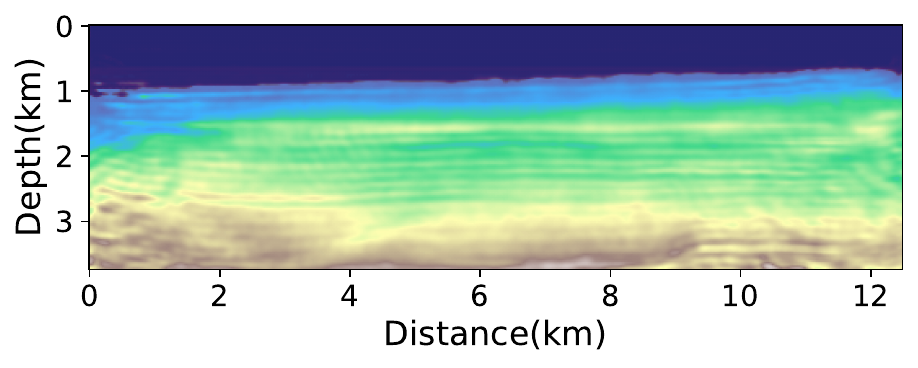}
    \includegraphics[width=0.48\textwidth]{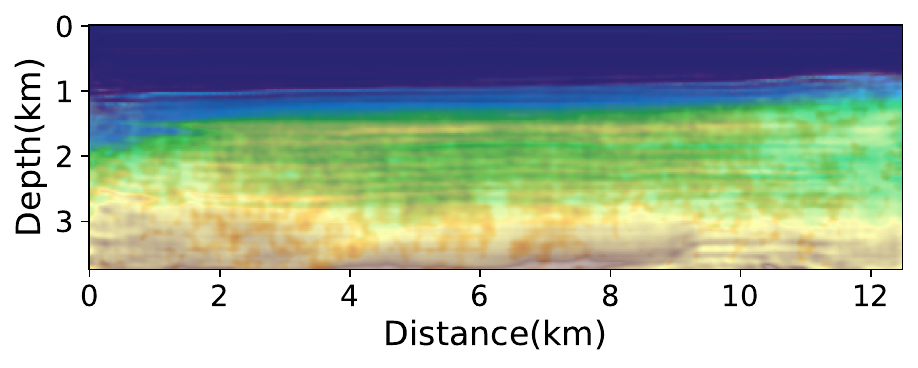}    
    \includegraphics[width=0.4\textwidth]{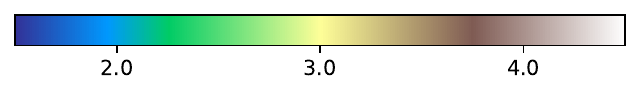}
    \includegraphics[width=0.4\textwidth]{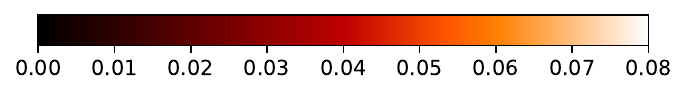}   
    \caption{The mean and standard deviation for the inverted models of field data at different iterations. Top row: The mean of the initial and Inverted velocity models. Middle: The standard deviation of the initial and inverted models. Bottom row: Overlay of initial and inverted means on their corresponding standard deviations.}
    \label{fig:14-lt}
\end{figure*}

Figures \ref{fig:14-lt} shows the mean and standard deviation of the posterior obtained through the application of the proposed conditional CNN network. The high standard deviation observed just below the water interface arises due to the uncertainty associated with the substantial transition (the location of the interface) from water to solid. This phenomenon is expected and serves to validate the credibility of the results. Additionally, it is noteworthy that higher uncertainty is seen in regions characterized by higher velocity, as the resolution decreases with increasing wavelength. Lastly, we observe that uncertainty below the water layer and after the first wavelength increases with depth.\\

\begin{figure*}[!tb]
    \centering
    \includegraphics[width=0.48\textwidth]{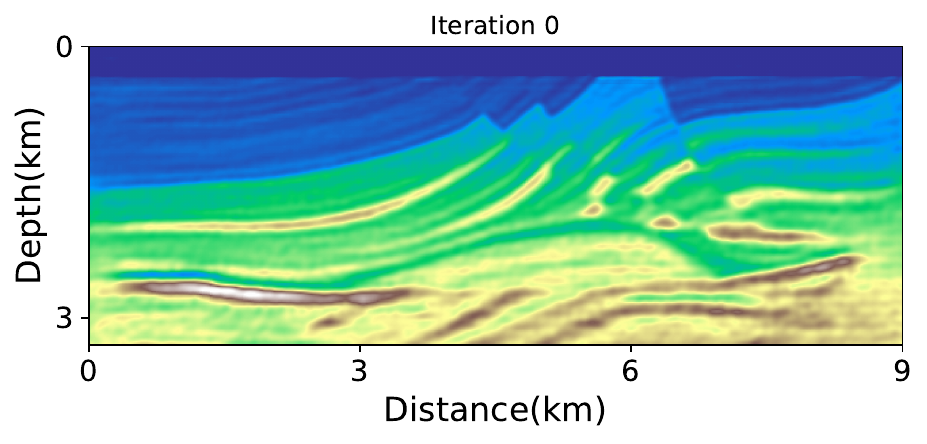}
    \includegraphics[width=0.48\textwidth]{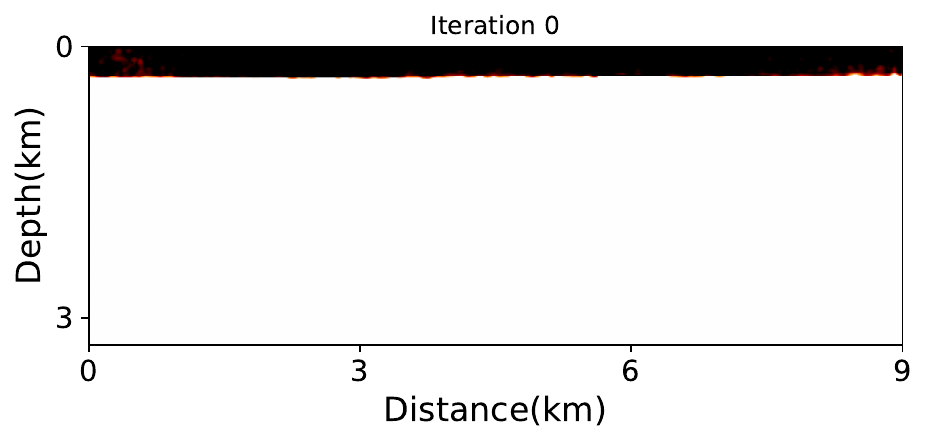}
    \includegraphics[width=0.48\textwidth]{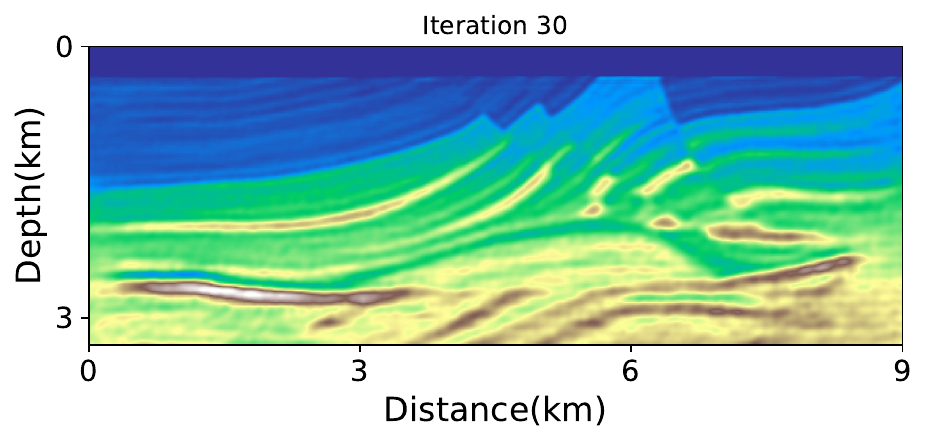}
    \includegraphics[width=0.48\textwidth]{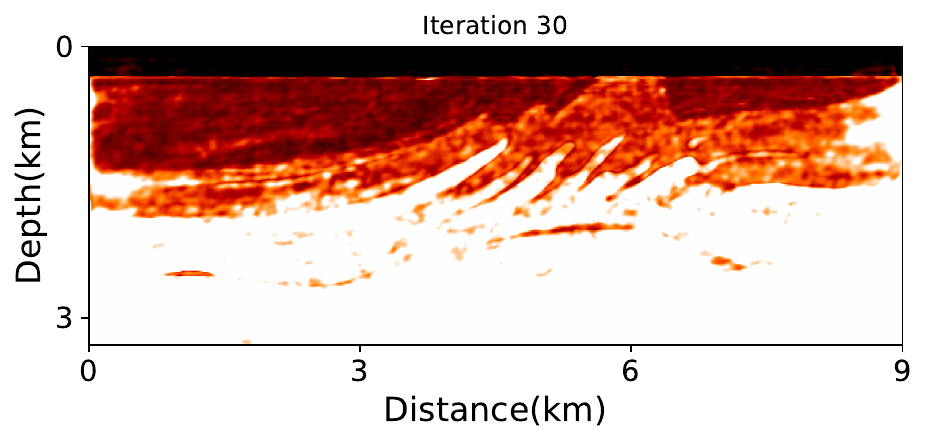}
    \includegraphics[width=0.48\textwidth]{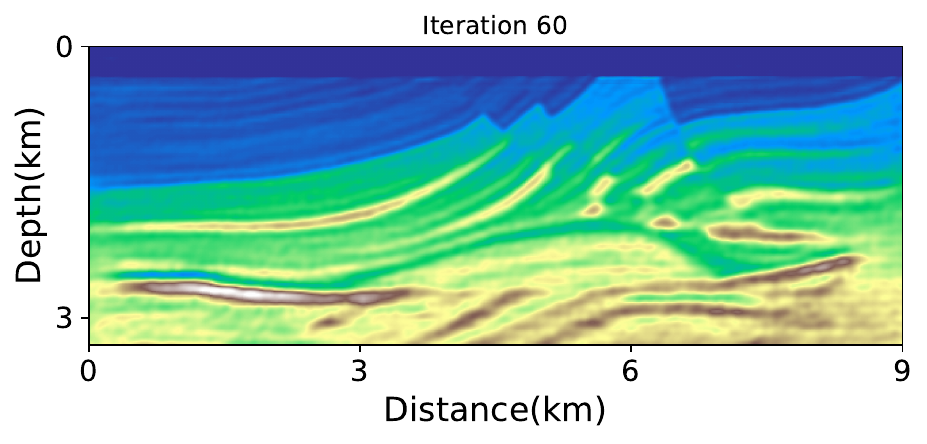}
    \includegraphics[width=0.48\textwidth]{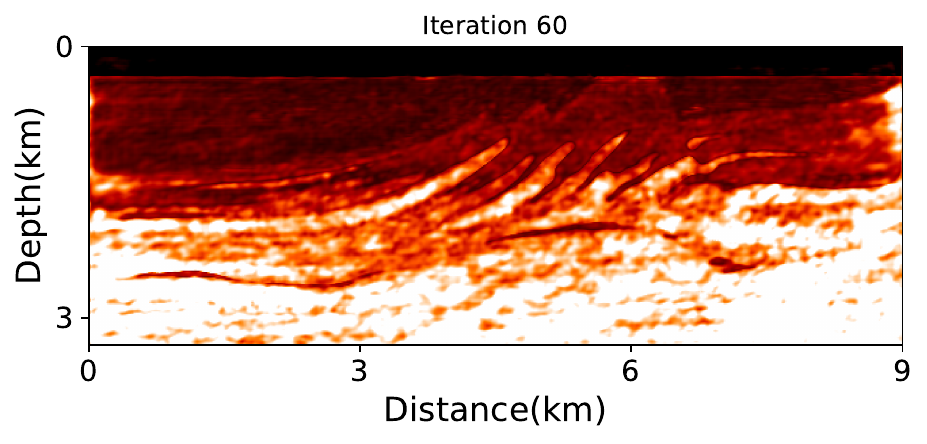}
    \includegraphics[width=0.48\textwidth]{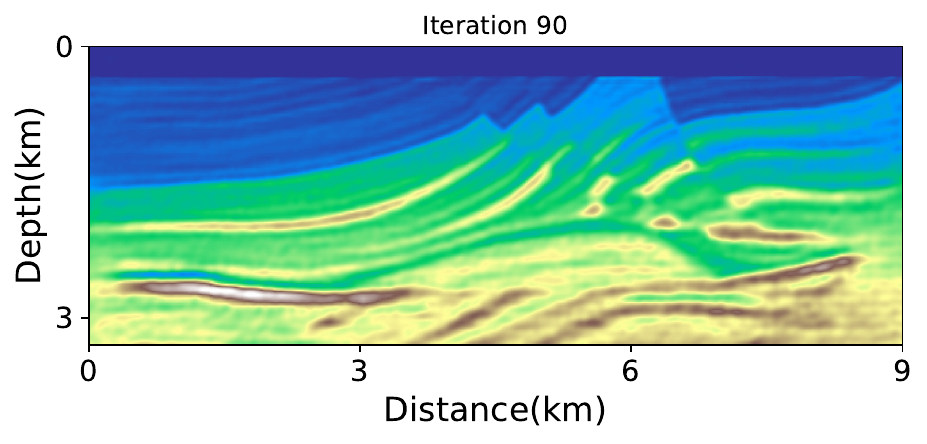}
    \includegraphics[width=0.48\textwidth]{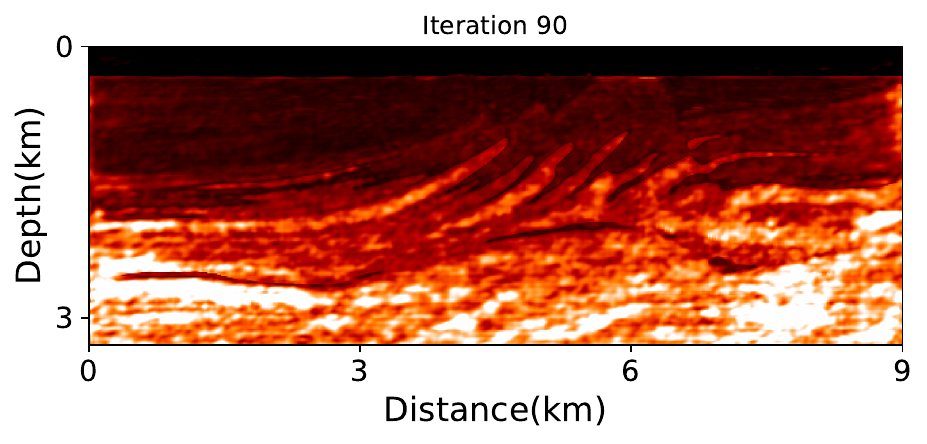}
    \includegraphics[width=0.48\textwidth]{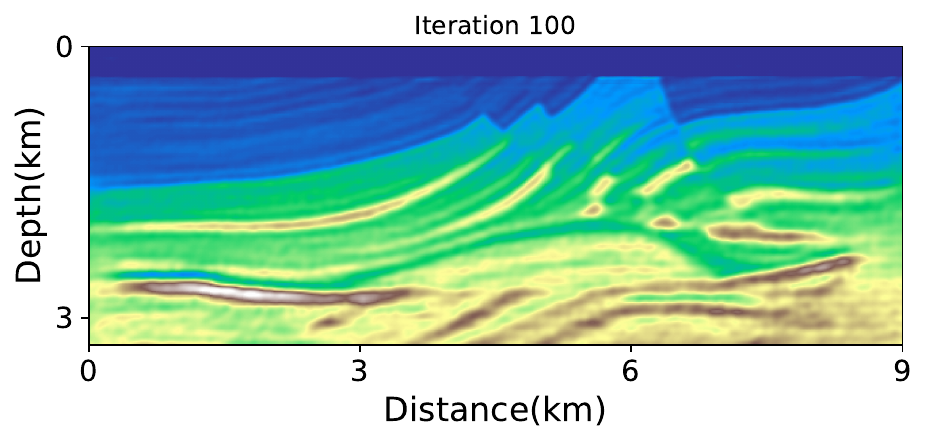}
    \includegraphics[width=0.48\textwidth]{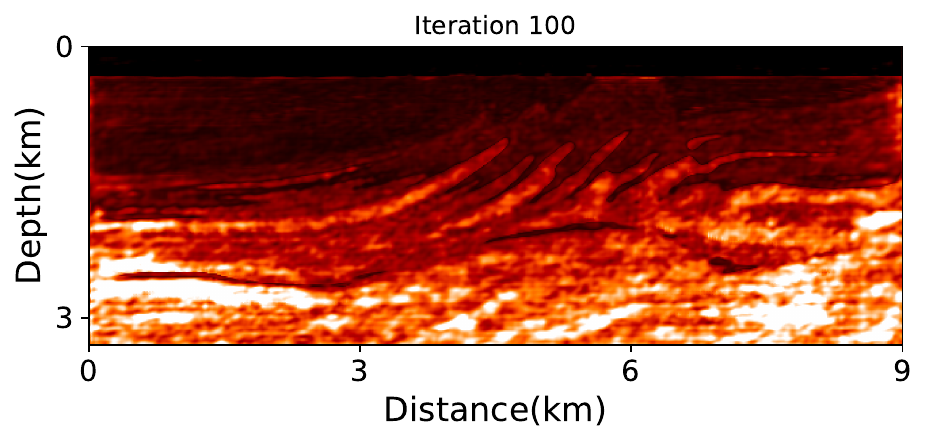}
    \centering
    \includegraphics[width=0.4\textwidth]{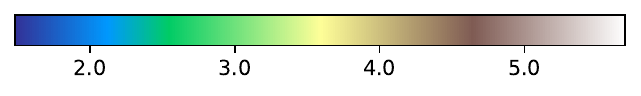}
    \includegraphics[width=0.4\textwidth]{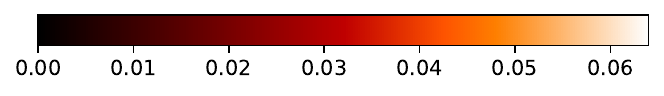}   
    \caption{Examples of the mean and standard deviation for the inverted models from the larger network at different iterations. }
    \label{fig:15-lt}
\end{figure*}

%% file: 05_Discussions.tex
\section{Discussion}
In this paper, we introduced a conditional CNN FWI framework, which enables the quantification of uncertainty in features captured by the neural network.  Expanding upon this conditional CNN network, we can sample infinite from the prior or posterior distribution, which can provide realizations for additional tests.

The high computational cost of FWI often complicates the evaluation of FWI uncertainties. In addition, most of the literature on uncertainty analysis mainly focuses on the scattering component, which ignores the uncertainty of the transmission component \cite[e.g.,][]{gebraad2020bayesian, zhang2021introduction, zhang20233}. Inspired by \cite{izzatullah2023frugal}, we use the Gaussian Random Field (GRF) perturbation to inject both scattering and wave propagation perturbations into the prior distribution. As a result, the uncertainty images shown in our paper include high uncertainty due to limited illumination, which can be observed on the sides of the model.

Additionally, we conducted a comparative assessment of our approach when using a neural network of a larger size, consisting of 8 million parameters. Figures \ref{fig:15-lt} shows the corresponding mean and standard deviation for the inverted models. Although the mean and standard deviation contain some additional features due to a change in its salient representation, the network requires additional time for convergence to better capture these features. The numerical experiments are performed on a Quadro RTX 8000 GPU using PyTorch \cite{paszke2017automatic} using the DeepWave package for FWI \cite{richardson_alan_2023}. Moreover, when comparing the computational time required by different CNN networks for 100 iterations with 50 particles of the Marmousi model, the smaller network took approximately 248 minutes, while the larger network took around 264 min. \\

%% file: 06_Conclusions.tex
\section{Conclusions}
We proposed a conditional Convolutional Neural Network (CNN) as an image prior to quantify FWI uncertainties. The condition is used to identify the samples of the prior distribution and later after FWI of the posterior distribution. We use Gaussian Random Fileds (GRF) perturbations to the initial or current velocity model to establish our prior samples. Subsequently, we integrate the pre-trained conditional CNN on these priors into the FWI process to generate samples from the posterior distribution. These samples can be utilized to quantify uncertainties in the features captured by conditional CNN, achieved by measuring the mean and standard deviation. The outcome obtained from the Marmousi model and field data demonstrate the effectiveness and efficiency of the proposed conditional CNN network.\\

%% file: 07_Acknowledgement.tex
\section*{Acknowledgment}
We thank KAUST for its support and the SWAG for collaborative environment. We also thank Dr. Shijun Cheng for the useful discussions. The authors thank the China Scholarship Council (grant nos. 202306450089) and DeepWave sponsors for supporting this research.